\documentclass[12pt]{iopart} % use for one-column layout

\usepackage{xcolor}
\usepackage{graphicx}
\usepackage{adjustbox}
\usepackage{placeins}
\usepackage[T1]{fontenc}
\usepackage{lipsum}
\usepackage{csquotes}
\usepackage{colortbl}
\usepackage{bm}
\eqnobysec
\usepackage[
backend=biber,
 style=chem-angew,
]{biblatex} 
\begin{document}

\pdfminorversion=4

\title[Ab-Initio Core-Hole Simulations]{The Nuts and Bolts of Ab-Initio Core-Hole Simulations for K-shell X-Ray Photoemission and Absorption Spectra}

\author{Benedikt P Klein$^{1,2}$, Samuel J Hall$^{1,3}$ and Reinhard J Maurer$^1$}
\address{$^1$Department of Chemistry, University of Warwick, Gibbet Hill Rd, Coventry, CV4 7AL, UK}
\address{$^2$Diamond Light Source, Harwell Science and Innovation Campus, Didcot, OX11 0DE, United Kingdom}
\address{$^3$MAS CDT, Senate House, University of Warwick, Gibbet Hill Rd, Coventry, CV4 7AL, UK }

\ead{r.maurer@warwick.ac.uk}
\vspace{10pt}
\begin{indented}
\item[]\today
\end{indented}

\begin{abstract}

X-ray photoemission (XPS) and Near Edge X-ray Absorption Fine Structure (NEXAFS) spectroscopy play an important role in investigating the structure and electronic structure of materials and surfaces. {\it{Ab-initio}} simulations provide crucial support for the interpretation of complex spectra containing overlapping signatures. Approximate core-hole simulation methods based on Density Functional Theory such as the Delta-Self-Consistent-Field ($\Delta$SCF) method or the transition potential (TP) method are widely used to predict K-shell XPS and NEXAFS signatures of organic molecules, inorganic materials and metal-organic interfaces at reliable accuracy and affordable computational cost. We present the numerical and technical details of our variants of the $\Delta$SCF and transition potential method (coined $\Delta$IP-TP) to simulate XPS and NEXAFS transitions. Using exemplary molecules in gas-phase, in bulk crystals, and at metal-organic interfaces, we systematically assess how practical simulation choices affect the stability and accuracy of simulations. These include the choice of exchange-correlation functional, basis set, the method of core-hole localization, and the use of periodic boundary conditions. We particularly focus on the choice of aperiodic or periodic description of systems and how spurious charge effects in periodic calculations affect the simulation outcomes. For the benefit of practitioners in the field, we discuss sensible default choices, limitations of the methods, and future prospects.

\end{abstract}

%\noindent{\it Keywords}: X-ray photoemission, X-ray absorption, XPS, XAS, NEXAFS, XANES, ELNES, density functional theory, delta self consistent field, transition potential
%\submitto{\JPCM}
\maketitle
%%%%%%%%%%%%%%%%%%%%%%%% here it is %%%%%%%%%%%%%%%%%%%%%%%%%%%%%%%%%
%%%%%%%%%%%%%%%%%%%%%%%% here it is %%%%%%%%%%%%%%%%%%%%%%%%%%%%%%%%%

\section{Introduction}
 
X-ray photoelectron spectroscopy (XPS) and X-ray absorption spectroscopy (XAS, often called near edge X-ray absorption fine structure, NEXAFS, or X-ray absorption near edge structure, XANES) play an important role in the characterization of materials and surfaces~\cite{Frati2020,Stoehr1992,SurfaceScienceTechniques2013}. XPS is routinely used to provide information on the chemical composition and the oxidation state of elements in various systems, including oxides~\cite{Bagus2019}, two-dimensional materials such as graphene~\cite{Presel2015,Presel2017}, and metal-organic interfaces~\cite{Netzer1992,Gao1999,Papp2013}. NEXAFS and, particularly, angular- or polarization-dependent NEXAFS can provide insight into the electronic structure and orientation of thin films~\cite{Breuer2015}, surfaces and adsorbates~\cite{Haase1985}, and even liquids~\cite{Niskanen2017}. Both methods are particularly suitable for surface characterization: XPS is inherently surface sensitive due to the small escape depth of the electrons from condensed matter and NEXAFS can make use of the same effect by choosing a suitable electron yield detection~\cite{Stoehr1992,Powell2009}.

\begin{figure}
    \centering
    \includegraphics[width=0.65\linewidth]{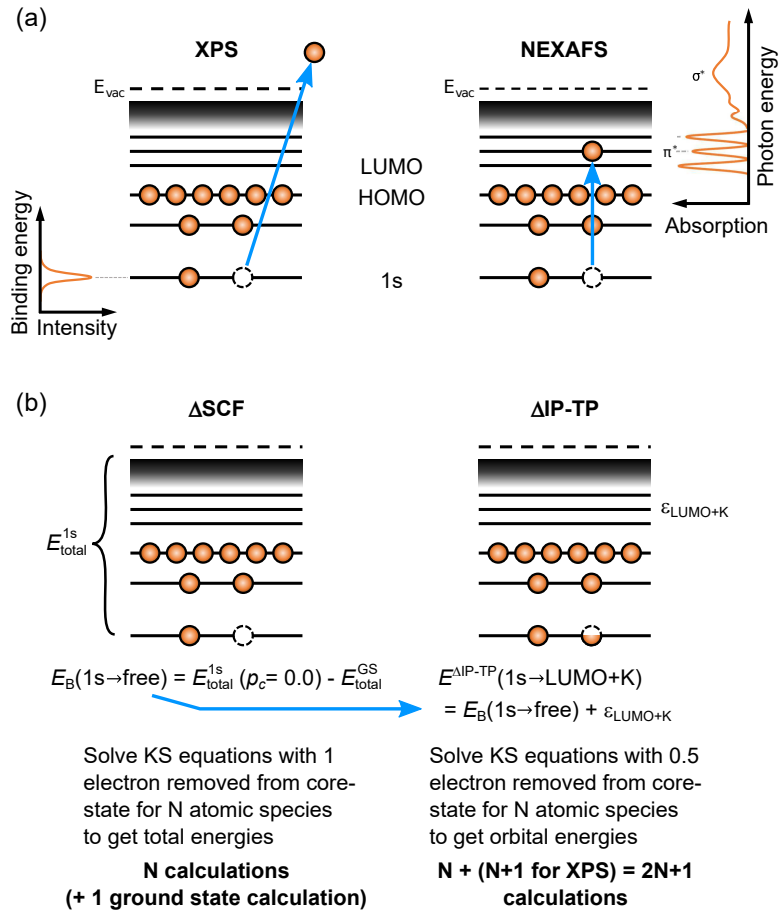}
    \centering\caption{\label{fig:intro}(a) Principles of X-Ray photoelectron spectroscopy (left) and near-edge X-ray absorption fine-structure (NEXAFS) spectroscopy (right). XPS detects photoelectrons emitted from core-levels after absorption of X-ray radiation. This provides information about the chemical environment of the core-states for each atom. NEXAFS probes secondary processes that are associated with the excitation of core-electrons into unoccupied valence states. This provides information about the unoccupied valence states of the system. (b) In this paper, we calculate XPS spectra with the Delta-Self-Consistent-Field ($\Delta$SCF) method (left) and NEXAFS spectra with the ionization-potential-corrected Transition Potential ($\Delta$IP-TP) method.}
    \label{fig:xps_nexafs}
\end{figure}

In core-level spectroscopy (see \fref{fig:intro}a), the sample is exposed to an X-ray beam, leading either to the emission of a photoelectron from a core-level (XPS) or the excitation of the core-electron to unoccupied electronic states (NEXAFS). For XPS, a fixed photon energy is used and the spectrum is acquired by scanning over the kinetic energy of the photoelectrons. For NEXAFS, the photon energy is varied and the energy-dependent absorption is measured. Ideally, the measured spectrum exhibits sharp signatures which can be directly assigned to originate from individual core-levels (XPS, \fref{fig:intro}a left) or from transitions between core-levels and unoccupied states (NEXAFS, \fref{fig:intro}a right). Changes in the chemical composition, molecular conformation or even orientation, or electronic structure of a sample will be reflected in the position, shape, and intensity of spectral signatures. Therefore both methods provide direct evidence of such changes.

However, the rich structure and overlapping features present in XPS and NEXAFS spectra often pose a challenge to the unambiguous assignment of all peaks. Here, the theoretical analysis and computational simulation of the electronic structure and transitions can contribute and produce additional insights. The first-principles-based simulation of core-level spectra is widely employed for this purpose. Established methods include the use of time-dependent Density Functional Theory (DFT)~\cite{Besley2020,Tenorio2018}, Many Body Perturbation Theory methods such as GW (for XPS) and Bethe-Salpeter equation calculations (for neutral excitations)~\cite{Golze2018,Golze2019}, as well as multiconfigurational wave function methods and Coupled Cluster calculations~\cite{Helmich-Paris2020,Liu2019,Vidal2020}. In addition, more cost-effective DFT methods that approximate the effects of an excited core-hole within a ground-state DFT formalism have been proposed. The latter methods include Delta-Self-Consistent-Field ($\Delta$SCF) DFT~\cite{Gunnarsson1976,VonBarth1979,VonBarth1980} to simulate XPS spectra and the transition potential (TP) method~\cite{Triguero1998} to simulate NEXAFS spectra. Many variations of these methods have been proposed in the past~\cite{Takahata2003,Prendergast2006,Kahk2019} with varying uptake in the community. These approximate methods are particularly suitable to study K-shell spectroscopy, i.e. the emission or excitation of 1s electrons, because higher lying states feature additional challenges such as spin-orbit coupling and delocalization over several atoms~\cite{Bunau2013,Donval2017}.

Approximate core-hole excitation methods to simulate core-level spectroscopy differ in how many electrons are excited, how the core-hole constraint is imposed and localized~\cite{Michelitsch2019}, which underlying basis set is used, e.g. if all-electron local atomic basis functions or pseudopotential-plane-wave approaches are used~\cite{Gao2009,Kahk2019,Ambroise2019}, and how relativistic effects are treated. The empirical wisdom behind simulating reliable and accurate core-hole spectra is often very software- and application-specific and is rarely the focus of studies, but rather relegated to supplemental information, even as it is crucial for the achieved performance.

In this paper, we present our approach to approximate core-hole simulations and discuss  methodological and technical intricacies. We analyze the dependence of simulated spectra with respect to all relevant simulation parameters to establish what considerations are required to produce reliable and converged XPS and NEXAFS spectra that are independent of numerical choices. We draw from our experience in simulating XPS and NEXAFS signatures of organic molecules adsorbed at metal interfaces~\cite{Blobner2015,Diller2017,Klein2019,Klein2019a,Klein2020} and highlight the challenges associated with simulating 1s XPS and K-edge NEXAFS spectra for organic molecules in gas-phase, in bulk crystals, and adsorbed at surfaces. After a brief introduction into the methodology and practical steps of XPS and NEXAFS calculations in \sref{methods-section}, we analyse how spectra depend on the choice of core-hole localization method (\sref{section-core-hole}), exchange-correlation functional (\sref{section-xc}), basis set (\sref{section-basis}), and model representation (cluster or periodic) before highlighting the challenges specific to bulk and surface calculations in \sref{section-bulk-surface}. On the latter topic, we specifically discuss the role of charging artefacts in periodic calculations. We further showcase the strength of analysing simulated NEXAFS spectra in terms of their angular and polarization dependence and by decomposing them into atomic and molecular orbital contributions to disentangle chemical shifts from final state screening effects (\sref{section-atom-MO}).

\section{Methods \label{methods-section}}

\subsection{Ab-Initio Simulation of X-Ray Photoemission Spectroscopy (XPS)}

%%%%%%%%%%%%%%%start with a basic definition of XPS experiment and binding energies

\paragraph{Principle of XPS:}
Since the first observation of the photoelectric effect by Hertz and Hallwachs, photoemission spectroscopy has become one of the most important techniques for the chemical analysis of molecules and materials~\cite{Reinert2005}. Upon exposure to X-ray radiation with frequency $\nu$, photoelectrons with kinetic energy $E_{\mathrm{kin}}$ and momentum $\mathbf{k}$ can be detected following the relationship:
\begin{equation}\label{eq:xps_eb}
    E_{\mathrm{kin},\mathbf{k}} = h\nu -E_\mathrm{B} - \Phi
\end{equation}
In \eref{eq:xps_eb}, $E_\mathrm{B}$ is the binding energy of the electron and $\Phi$ is the work function of the material. A typical photoelectron spectrum plots the intensity of an electron detector signal against the binding energy (from highest to lowest binding energy). Depending on the used photon energy, the intensity~$I$ of the photoelectrons with kinetic energy $E_{\mathrm{kin},\mathbf{k}}$ is proportional to the photoelectron cross section $\sigma_{\mathbf{k}}$ as follows~\cite{Heinz2012}:
\begin{equation}\label{eq:xps_i}
    I_{\mathbf{k}}(h\nu) \propto n  \,\lambda_{in} A \,J_{h\nu}\,\sigma_{\mathbf{k}}(h\nu)
\end{equation}
where $\lambda_{in}$ is the inelastic mean free path (IMFP), $n$ is the number of atoms of this species per volume, $A$ is the surface area covered by the X-ray spot, and $J_{h\nu}$ is the flux density of incoming photons of frequency $\nu$. With few exceptions~\cite{Diller2014,Taucher2016}, the IMFP is mostly neglected in simulations as, within the typical kinetic energy range (a few hundred electronvolts up to 1 keV), the IMFP of photoelectrons is of the order of 1~nm, which amounts to few atomic layers from the surface~\cite{Powell2009}. The cross section $\sigma_{\mathbf{k}}$ itself can be related to the spectral function $A_{ii}(E_{\mathrm{kin,\mathbf{k}}}-h\nu)$ of photoemission of an electron with momentum $\mathbf{k}$ from core-state $i$  via Fermi's golden rule and the sudden approximation~\cite{Hedin1999,Golze2019}:
\begin{equation}\label{eq:xps_spectral}
    \sigma_{\mathbf{k}}(h\nu) = \sum_{i} \left|\langle\psi_{\mathbf{k}}|\mathbf{p}\cdot\mathbf{\mathbf{A}(\mathbf{r})}|\psi_i \rangle \right|^2 A_{ii}(E_{\mathrm{kin,\mathbf{k}}}-h\nu)
\end{equation}
where $\mathbf{A}(\mathbf{r})$ is the vector potential of the incident electromagnetic field and $\mathbf{p}$ is the momentum operator. 

Within many-body theory, the spectral function arises from the single particle Green's function that describes the removal or addition of particles to the system~\cite{Golze2019} and describes the frequency-dependent probability of photoemission and the associated lineshape. The positions of the poles in the Green's function correspond to the electron addition/removal energies with respect to the Fermi energy, i.e. the binding energies of the associated quasi-particles. This takes account of all relevant interaction processes, including collective screening, electron-electron scattering and relaxation effects due to the introduction of the core-hole, all of which arise from many-body correlation. For example on metals, XPS spectra typically show a characteristic asymmetry, which stems from the interaction of the photoelectron with electron-hole-pair excitations close to the Fermi level~\cite{Doniach1970}. Full many-body calculations can provide very accurate XPS spectra that fully account for many spectral features~\cite{Golze2018,Golze2019}. Such calculations are, however, associated with high computational cost and are all but unfeasible for systems larger than several tens of atoms, and therefore remain unpractical for organic crystals or metal-organic interfaces. 

%%%%%%%%%%%%%%%%%%%%%%  Approximate DFT-based methods to simulate binding energies

In practice, the main features of XPS spectra are often simulated by effective single particle approaches. Instead of evaluating the cross section from the full spectral function, it is instead approximated by a set of single particle transitions between effective Kohn-Sham states within Density Functional Theory. By assuming independent electrons [$A_{ij}(h\nu)=\delta_{ij}\delta(h\nu-\varepsilon_i)$], we can simplify \eref{eq:xps_spectral} to the well known Fermi's golden rule expression~\cite{Green2005}:
\begin{equation}\label{eq:xps_fermi}
\sigma_{\mathbf{k}}(h\nu) = \sum_{i} \left|\langle\psi_{\mathbf{k}}|\mathbf{p}\cdot\mathbf{A}(\mathbf{r})|\psi_i \rangle \right|^2 \delta(E_{\mathrm{kin,\mathbf{k}}}-h\nu-\varepsilon_i)
\end{equation}
If no selection rules are to be considered, for example if no angular dependence is required (as would be in angular resolved photoemission spectroscopy~\cite{Puschnig2011, Shang2011}), the matrix elements in \eref{eq:xps_fermi} are often neglected as all final states are assumed to be extended plane wave states. This amounts to associating each excitation that originates from an atom-centred core-state $i$ with the same intensity, i.e. eliminating all but $n$ and $\sigma_{\mathbf{k}}(h\nu)$ from \eref{eq:xps_i}. The result of such a calculation is a spectrum with $N$ infinitely narrow $\delta$-peaks where $N$ is the number of atoms, each with the same intensity and located at the energetic position of the binding energy $E_\mathrm{B}$ of the respective atom. 

The line shape of a calculated XPS signature is then often approximated by assuming a Gaussian, Lorentzian, or convoluted Gaussian-Lorentzian (i.e., Voigt) function for each atomic peak. The summation of all peak functions leads to a spectral shape comparable to what is commonly observed in experiments. The Lorentzian broadening is justified with life-time effects of the excited states, while Gaussian broadening simulates experimental effects (e.g., analyser resolution). A simplified way to incorporate both kinds of broadening into the simulated spectra, which was chosen for this publication, is the use of a pseudo-Voigt function, which reduces the convolution needed for a proper Voigt function to a summation operation~\cite{Schmid2014,Schmid2014cor}. When using such functions, the parameters for line-width (full width at half maximum, FWHM) and Gaussian-Lorentzian ratio of the broadening are chosen according to the expected or observed experimental resolution. Using these approximations to obtain intensities and line shapes, the problem of simulating XPS spectra is reduced to finding the binding energies $E_\mathrm{B}$ of the relevant electronic core-states. 

%Discuss how XPS based on effective KS states neglects core-hole effects
\paragraph{Ab-initio calculation of core-level binding energies:}

Many different approaches exist to calculate core-level binding energies including Coupled Cluster (CC)~\cite{Zheng2019},  Equation-of-Motion CC (EOM-CC)~\cite{Vidal2020}, the GW method~\cite{Golze2018,Golze2019}, and real-time TD-DFT~\cite{Kas2015}. The simplest approach, however, is the use of Koopman's theorem \eref{eq:koopmans}, which relates the ionization potential of an electronic state in Hartree-Fock theory with the negative of the relevant Hartree-Fock eigenstate~\cite{Koopmans1934}. Despite the fact that in Density Functional Theory (DFT) this relationship only holds for the highest occupied Kohn-Sham state~\cite{Janak1978,Bellafont2015}, Kohn-Sham (KS) energies of core-levels are often used to estimate the chemical shift contribution to the binding energy, i.e. the displacement of the core-level before removal of the electron (also termed initial state effect)~\cite{Williams1978,PueyoBellafont2015,Chong2002}.
\begin{equation}\label{eq:koopmans}
    E_\mathrm{B}(i) = -\varepsilon_{i,\mathrm{HF/KS}}
\end{equation}

%Discuss the role of core-hole relaxation and how it can be incorporated into approximate XPS simulations with the DeltaSCF method.

In most cases, however, core-hole relaxation effects (or final state effects) cannot be neglected in the prediction of binding energies. The Delta-Self-Consistent-Field DFT ($\Delta$SCF-DFT or $\Delta$SCF) method~\cite{Slater1972a,Gunnarsson1976,VonBarth1979}, which includes relaxation effects, is therefore the most common approach to simulate core-level binding energies~\cite{PueyoBellafont2015,Bagus1965,Johansson1999,Besley2009,Vines2018}. As shown in \fref{fig:intro}b, left, the $\Delta$SCF method calculates binding energies as energy differences between two self-consistent Kohn-Sham DFT calculations, namely the ground-state calculation and the core-hole excited calculation, where the electron population $p_c$ of the 1s core-state has been constrained to remove one electron:
\begin{equation}\label{eq:deltascf}
    E_{\mathrm{B}}(1\mathrm{s}\rightarrow \mathrm{free}) = E_{\mathrm{total}} ^{1s}(p_c=0.0)  - E_{\mathrm{total}}^{\mathrm{GS}}
\end{equation}
Other methods have been proposed that introduce core-holes equivalent to half an electron (so-called Slater transition state approach)~\cite{Slater1972a}, or two-thirds of an electron~\cite{Williams1975}, although full removal of an electron has been shown to provide the best results~\cite{Triguero1998}. Unfortunately, often a simple change of orbital occupation does not provide a sufficient constraint to localize the core-hole onto the atom-centred core-level~\cite{Chong2007,Noodleman1982}. This is particularly true if the molecule contains symmetry-equivalent atoms with degenerate core-states. Many different approaches have been proposed in the past to overcome this, such as the maximum overlap method (MOM)~\cite{Besley2009}, using a localized hole in a frozen core-state~\cite{Takahata2003}, shifting results between symmetric MOs~\cite{Chong2010}, and localization via initialization with a density calculated for an increased atomic core-charge~\cite{Kahk2019}. Due to those difficulties, we will discuss the issue of core-hole localization in detail in \sref{section-core-hole}.

An additional problem for calculations of systems with periodic boundary conditions is the treatment of the charge introduced by the core-hole calculations. The unit cell of every periodic calculation has to be net-neutral to avoid  failure of the Ewald summation. Therefore, a uniform negative charge background is introduced into the calculation of the positively charged unit cell with the core-hole, which might lead to unpredictable changes in the calculated energies. This charging issue in periodic systems has previously been recognized as problematic~\cite{Okazaki2004,Taucher2020}. The effect of the background charge can in principle be reduced by increasing the unit cell size. Unfortunately, the necessary sizes to reduce the introduced shift to a negligible size are commonly thought to be too large for practical uses. However, while we show in \sref{section-basis} that the absolute energies are not easily converged, we also found that the relative shifts are quite robust with varying cell size.

\subsection{Ab-Initio Simulation of NEXAFS}

%%%%%%%%%%%REVIEW OF LITERATURE
\paragraph{Principles of NEXAFS: }

Excellent accounts of the measurement and first-principles simulation of X-ray absorption have been given previously by H\"ahner~\cite{Hahner2006}, Frati \emph{et al.}~\cite{Frati2020}, Tanaka and Mizoguchi~\cite{Tanaka2009,Mizoguchi2009,Mizoguchi2010}, Norman and Dreuw~\cite{Norman2018}, and Guda \emph{et al.}~\cite{Guda2019}. We do not attempt to cover the full breadth of literature on the subject here and will focus mostly on simulation with  approximate DFT methods.

%%%%%%%%%%%%%%%%%%%%% BASICS OF NEXAFS

An X-ray absorption spectrum is experimentally recorded by varying the photon energy of incident electromagnetic radiation and measuring a detector signal, which is proportional to the absorption of X-rays by the sample, and therefore also proportional to the photon energy dependent cross section $\sigma(h\nu)$. This X-ray absorption cross section originating from excitation of core-state $i$ as given by Fermi's golden rule in the single particle and dipole approximation is~\cite{Stoehr1992}:
 \begin{equation}\label{eq:NEXAFS}
     \sigma_i(h\nu) \propto  \sum_f  \left|\langle\psi_f|\mathbf{p}\cdot\mathbf{A}(\mathbf{r})|\psi_i \rangle \right|^2 \delta(h\nu+E_i-E_f)
 \end{equation}
 where $\psi_f$ and $\psi_i$ correspond to the wave functions of the final valence  and the initial core electronic states, respectively. $\mathbf{p}$ is the dipole  operator and $\mathbf{A}(\mathbf{r})$ is the vector potential of the incident electromagnetic field. This can be further simplified by assuming a classical polarization wave with polarization along a unit vector $\mathbf{e}$, leading to $\mathbf{A}(\mathbf{r})=\mathbf{e}A_0e^{i\mathbf{k}\cdot\mathbf{r}}$~\cite{Rehr2000}. For local core-excitations, the dipole approximation is valid ($e^{i\mathbf{k}\cdot\mathbf{r}}=1$) and yields:
 \begin{equation}\label{eq:NEXAFS2}
\sigma_i(h\nu) \propto \sum_f \left|\langle\psi_f|\mathbf{e}\cdot\mathbf{p}|\psi_i \rangle \right|^2 \delta(h\nu - (\underbrace{E_f-E_i}_{\Delta E_{fi}}) )
 \end{equation}
Therefore, the intensity of a NEXAFS spectrum is defined by the nature and overlap of initial and final states. The $\delta$-function ensures energy conservation between the energy of the incident light and the transition energy $\Delta E_{fi}$. To obtain the spectral shape, the discrete $\delta$-functions of each excitation are typically replaced by a model of the finite line shape to account for the life-time of final states and other broadening effects, in analogy to the XPS line shapes (\emph{vide infra}).

%%%%%%%%%%%%%%%%%%%%%%%%%%%%%% Many methods exist, we will focus on Transition potential

\paragraph{Simulation of NEXAFS spectra: } 
A large variety of methods to simulate X-ray absorption spectra have been proposed including linear response based on CASSCF calculations~\cite{Helmich-Paris2020}, TD-DFT~\cite{Lopata2012,Tenorio2018}, and Bethe-Salpeter equation calculations~\cite{Rehr2005,Olovsson2009,Vinson2011,Gilmore2015}. Here we discuss NEXAFS in the context of molecular bulk and metal-organic surface systems, where the choice of method is much more limited due to the intrinsic size of and computational cost associated with such systems.  

A variety of more approximate DFT-based methods to calculate NEXAFS transition energies $\Delta E_{fi}$ have sprung from the $\Delta$SCF method. These include methods that explicitly simulate each transition between the core-hole and all unoccupied states, such as the $\Delta$SCF method and the Slater-Janak transition-state method~\cite{Slater1972a}, where one or half an electron are excited from the core-level $p_c$ to the relevant valence state $p_v$, respectively. The transition energy is given by the difference in total energy of the core-excited calculation $(p_c=0.0,p_v=1.0)$ and the DFT ground state $(p_c=1.0,p_v=0.0)$:
\begin{equation}
\Delta E^{\Delta \mathrm{SCF}} = E(p_c=0.0,p_v=1.0) - E(p_c=1.0,p_v=0.0)
\end{equation}
In this case, $M$ calculations are required, where $M$ is the number of valence states $v$ that will be considered. This procedure needs to be repeated for all relevant core-levels $N$, on which the core-hole is to be localized. For K-shell excitations, $N$ equals the number of symmetry-inequivalent atoms. The total number of DFT calculations for a single NEXAFS spectrum is therefore $(N\cdot M)+1$ (including the ground-state). Using this approach, the sheer amount of calculations can be prohibitive if the system contains a large number of atoms.

The transition potential (TP) method~\cite{Stener1995,Hu1996}, is an approximation based on Slater's transition state approach~\cite{Slater1972a}, which circumvents such a large number of calculations. Rather than calculating each excitation explicitly as a difference between the total energy of an excited state and the total ground-state DFT energy, the transition energy is defined as the difference between the Kohn-Sham energies of valence ($\varepsilon_v$) and core-states ($\varepsilon_c$). These KS energies are calculated for a system, where half an electron is removed from the core-state to model the effect of the core-hole on the transition:
\begin{equation}\label{eq:tp-method}
    \Delta E^{\mathrm{TP}} = \varepsilon_v(p_c=0.5,p_v=0.0)-\varepsilon_c(p_c=0.5,p_v=0.0)
\end{equation}
A single TP DFT calculation provides all KS states and therefore yields all NEXAFS transitions and the number of calculations for a spectrum is reduced to the number of atomic species $N$.
A variant of the TP method has been proposed that introduces a full core-hole (FCH), instead of a half core-hole~\cite{Hetenyi2004,Tanaka1999}:
\begin{equation}\label{eq:fch-method}
    \Delta E^{\mathrm{FCH}} = \varepsilon_v(p_c=0.0,p_v=0.0)-\varepsilon_c(p_c=0.0,p_v=0.0)
\end{equation}
Note that in both cases the (partial) electron removed from the core-state is not placed into the valence states and the system is not net neutral. The lack of charge neutrality, as is also the case for $\Delta$SCF calculations of XPS binding energies, can provide a challenge for simulations of bulk and surface systems due to the prospect of charge artefacts within periodic boundary conditions~\cite{Taucher2020}. Several solutions to this problem have been proposed~\cite{Ozaki2017}, including variants of TP that restore charge neutrality by placing the removed charge into the lowest unoccupied molecular orbital (LUMO) of the system such as the XCH method which introduces a full core-hole~\cite{Prendergast2006,Michelitsch2019}: 
\begin{equation}\label{eq:xch-method}
    \Delta E^{\mathrm{XCH}} = \varepsilon_v(p_c=0.0,p_{\mathrm{LUMO}}=1.0)-\varepsilon_c(p_c=0.0,p_{\mathrm{LUMO}}=1.0)
\end{equation}
A comprehensive summary of different approximate core-hole constraint methods has recently been given by Michelitsch and Reuter, where the authors also introduced other previously not considered variants of the TP method~\cite{Michelitsch2019}. For a small set of organic molecules in gas-phase, the authors compare these different approaches and find that carbon and nitrogen K-edge NEXAFS spectra (intensities and excitation energies) are best represented by the XCH and XTP methods. The latter is a variation of TP where the removed half core-hole is neutralized with half an electron placed into the LUMO:
\begin{equation}\label{eq:xtp-method}
    \Delta E^{\mathrm{XTP}} = \varepsilon_v(p_c=0.5,p_{\mathrm{LUMO}}=0.5)-\varepsilon_c(p_c=0.5,p_{\mathrm{LUMO}}=0.5)
\end{equation}

While the TP and XTP methods seem to provide reliable and robust predictions of the relative peak positions and intensities in NEXAFS spectra~\cite{Michelitsch2019}, the XCH and $\Delta$SCF methods, which both introduce a full core-hole, seem to provide a better description of absolute energies~\cite{Triguero1998}. Whereas the half core-hole constraint in TP and XTP appears to accurately account for core-hole relaxation effects in valence states, the chemical shift and core-hole screening is more accurately represented with a full core-hole in $\Delta$SCF and XCH. This conundrum has previously already been recognized by Triguero \emph{et al.}~\cite{Triguero1998}. They proposed a hybrid approach, where the transition potential (half core-hole) is corrected by the ionization potential of the core state  calculated with a full core-hole. Following Triguero \emph{et al.}~\cite{Triguero1998}, we call this ionization-potential-corrected TP method $\Delta$IP-TP to distinguish it from the original TP method~\cite{Stener1995,Hu1996}. The energy of a transition between a core-level and an unoccupied valence state is calculated in $\Delta$IP-TP by combining the results of a $\Delta$SCF and a TP calculation (see also \fref{fig:intro}b, right):
\begin{equation}\label{eq:DeltaIP-TP}
\Delta E^{\mathrm{\Delta IP-TP}} = \varepsilon_v(p_c=0.5,p_v=0.0)+E_{\mathrm{B}}(1\mathrm{s}\rightarrow \mathrm{free})
\end{equation}
In \eref{eq:DeltaIP-TP}, $\varepsilon_v$ are the Kohn-Sham energies of the unoccupied states from the half core-hole TP calculation, whereas  $E_{\mathrm{B}}(1\mathrm{s}\rightarrow \mathrm{free})$ is the total binding energy of the core-level from a full core-hole $\Delta$SCF calculation as laid out in \eref{eq:deltascf}. This approach retains the accurate description of relative transition energies for each atomic center, but additionally improves not only the description of the chemical shift of each atom but also the absolute transition energies. We have employed such a hybrid approach in all our previous work on metal-organic interfaces and organic molecules, where we have achieved good  agreement with experimental data~\cite{Diller2017,Klein2019,Klein2019a,Klein2020}. In principle this method can also be employed on the basis of XTP instead of TP to calculate the KS-energies of the unoccupied states, as we will show below.

TP core-hole calculations are regularly performed in codes that use all-electron atomic orbital basis sets~\cite{Michelitsch2019} or codes that use plane waves and pseudopotentials~\cite{Gao2009}. In the former case, the binding energy from $\Delta$SCF can yield highly accurate absolute energies if well-designed basis sets with sufficiently flexible core-functions are used~\cite{Besley2009,Kahk2019,Ambroise2019,Sarangi2020}. In the case of pseudopotential plane-wave codes, in order to calculate dipole matrix elements, the full all-electron wave function needs to be reconstructed using the projector-augmented-wave method~\cite{Gao2009,Mizoguchi2009}. The frozen core adds to already existing errors that stem from the self-interaction error of approximate exchange-correlation functionals and the resulting absolute binding energies are typically not close to experiment. However, the application of a rigid global shift can still allow a fruitful comparison with the experiment.

\begin{figure*}
\centering
\includegraphics[width=\linewidth]{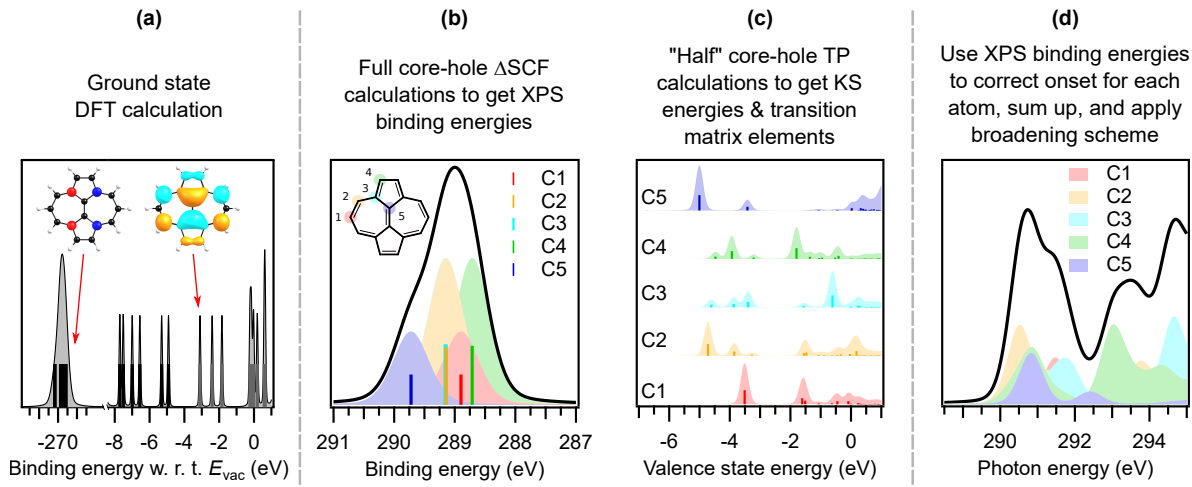}
\caption{Computational workflow to generate a K-shell NEXAFS spectrum with the $\Delta$IP-TP approach. (a) Ground state DFT calculation. (b) $N$ full core-hole $\Delta$SCF calculations to establish the chemical shifts of each atomic species. These calculations also provide the basis for the XPS spectrum. (c) $N$ half core-hole TP calculations to obtain the absorption spectra for each species. (d) Post-processing: Each atomic species is shifted according to its $\Delta$SCF binding energy and finally all atomic contributions are summed up. Additionally a suitable broadening scheme is included. The inset in (a) shows two exemplary core and valence states. The inset in (b) contains the  structural formula of our example molecule azupyrene and the peak assignment for the inequivalent carbon atoms.}
\label{fig:workflow}
\end{figure*}

\Fref{fig:workflow} summarizes the practical steps involved in performing a $\Delta$IP-TP (or $\Delta$IP-XTP) simulation of a K-edge NEXAFS spectrum: A ground-state DFT calculation (\fref{fig:workflow}a) is followed by $N$ $\Delta$SCF calculations (\fref{fig:workflow}b) with a full core-hole to determine the binding energy for each inequivalent core-level [$E_{\mathrm{B}}(1\mathrm{s}\rightarrow \mathrm{free})$ in \eref{eq:DeltaIP-TP}] and $N$ TP calculations (\fref{fig:workflow}c) with a half core-hole to determine the KS-energies of the unoccupied states [$\varepsilon_v$ in \eref{eq:DeltaIP-TP}]. These calculations each have to be performed $N$ times, because the full/half core-hole has to be introduced into each of the $N$ symmetry-inequivalent 1s states in the system. Each TP calculation first provides a self-consistently converged density and then works to accurately converge the unoccupied valence states. This approach is necessary to achieve the same convergence accuracy for the unoccupied states as for the occupied ones. Therefore, care needs to be taken that convergence thresholds are set appropriately for the former as their Kohn-Sham energies directly control the transition energies and the respective transition dipole matrix elements with the core-state.

Finally, the $\Delta$SCF binding energies are added to the TP eigenvalue spectra to align the contributions originating from different atomic species onto a single energy axis and the spectra are summed up (\fref{fig:workflow}d). The result is the final \emph{ab-initio} predicted spectrum. This spectrum can be plotted with the contributions from each core-state  (as shown) or further analysed by projecting to the contributions from valence states or molecular orbitals (see \sref{section-atom-MO}).

Both XPS and NEXAFS spectra are broadened to simulate typical experimental resolutions and ease the comparison to measured data. We chose to use a bare-bone scheme employing a minimal number of parameters and a mathematically simple implementation. While physically  more meaningful peak functions exist~\cite{Stoehr1992}, their proper application would be much more complex. The scheme used in this publication is still able to produce good agreement with experimental spectra and is superior for most practical purposes due to its simplicity of use. For the XPS spectrum, a fixed peak width and Gaussian/Lorentzian (G/L) ratio can normally be chosen for all peaks stemming from the same core-level. The spectrum is yielded as a summation over the pseudo-Voigt-functions~\cite{Schmid2014,Schmid2014cor} for each symmetry equivalent atom weighted by their abundance in the system. While the broadening function can also be chosen to be the same for the NEXAFS transitions of all atomic species, it has to change with photon energy in the NEXAFS spectrum. This approach is necessary to account for the difference in lifetime of states below and above the ionization threshold. Therefore, two different broadening schemes are employed for low and high photon energies. In the high photon energy regime, the width is larger and the G/L ratio is turned more Lorentzian, while the reverse is true for the low photon energy regime. Both regimes are linked by a linear gradient for width and G/L ratio, which crosses the ionization threshold, in line with common approaches in literature~\cite{Kolczewski2001,Puettner2004}. Systematic errors in the method and the self-interaction errors in the exchange-correlation functional make it furthermore necessary to globally shift the simulated spectra to align with the experimental energy axis both in the XPS and NEXAFS spectrum.

%%%%%%%%%%%%%%%%% orientation and angle dependence

\paragraph{Angular- and Polarization-resolved NEXAFS simulations:}

\begin{figure}
\centering
\includegraphics[height=8cm]{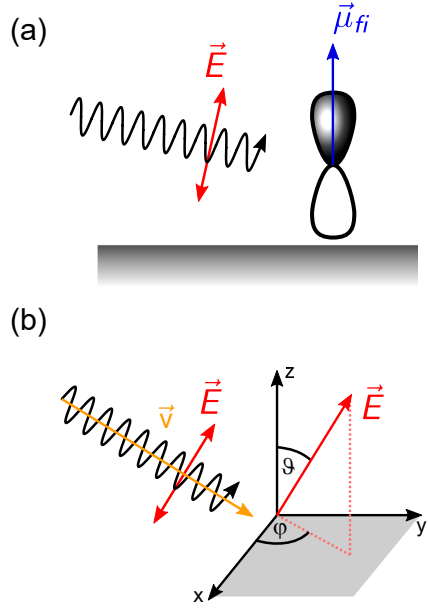}
\centering\caption{Schematic representation of the angular dependence in NEXAFS. (a) The intensity of a NEXAFS transition depends on the relative orientation of transition dipole moment $\boldsymbol{\mu}_{fi}$ and electrical field vector $\mathbf{E}$ (=$A_0 \cdot \mathbf{e}$ in the dipole approximation). Note that $\boldsymbol{\mu}_{fi}$ is fixed relative to the axes by the orientation of the calculated system. (For K-shell excitations, the $\boldsymbol{\mu}_{fi}$ orientation is also only determined by the final state). (b) The orientation of the electrical field vector can be described by two angles, the polar angle~$\vartheta$ and the azimuth~$\varphi$, $\mathbf{v}$ is the propagation direction of the radiation.}
\label{fig:angular_NEXAFS} 
\end{figure}

NEXAFS  can not only probe the character of electronic states, but also their orientation with respect to the incident light polarization~\cite{Hebert-Souche2000}. As can be seen in \eref{eq:NEXAFS}, the intensity depends on the orientational alignment of the direction of polarization of incident light $\mathbf{e}$ with respect to the vectorial change in dipole moment due to the electronic transition $\boldsymbol{\mu}_{fi}$. Assuming linear polarized light, we can simplify the matrix elements in \eref{eq:NEXAFS} as follows:
\begin{equation}
   \boldsymbol{\mu}_{fi} = |\langle\psi_f|\mathbf{e}\cdot\mathbf{p}|\psi_i \rangle |^2 \approx |\mathbf{e}\cdot \underbrace{\langle\psi_f|\mathbf{p}|\psi_i \rangle}_{\mathbf{D}_{fi}} |^2 =  \left(\sum_{r=x,y,z} e_r \cdot D_{fi,r}\right)^2
\end{equation}
Herein, $\mathbf{D}_{fi}$ are the dipole matrix elements between initial and final states $i$ and $f$, respectively. Therefore, the intensity is largest if the direction of the electric field vector $\mathbf{e}$ and the transition dipole vector are parallel. In case of a spherical 1s core-state, this means that the intensity is largest if the light polarization aligns with the electronic dipole of the final electronic state, see \fref{fig:angular_NEXAFS}a. 

For some systems, all orientational dependence of the NEXAFS transitions is likely to be averaged out in the experimental data due to random orientation within the  sample, for example in the gas phase, in liquids, or in amorphous or disordered polycrystalline solids. If this is the case for the experimental data we want to compare our simulations to, the orientational dependence can also be neglected in the calculations:
 \begin{equation}\label{eq:angle-averaged-NEXAFS}
     \mu_{fi}^{\mathrm{total}} = D_{fi,x}^2+D_{fi,y}^2+D_{fi,z}^2
 \end{equation}

For molecules adsorbed on metal surfaces, however, the polarization dependence of the NEXAFS signal (the so-called dichroism) is often present in the experiment and quite useful, because it enables the extraction of structural information about the studied system~\cite{Breuer2015}.
To describe the polarization dependence, we can define the polarization direction $\mathbf{e}=(e_x,e_y,e_z)$ of the incident light in terms of polar~$\vartheta$ and azimuthal~$\varphi$ angles with respect to the surface (see \fref{fig:angular_NEXAFS}b):
\begin{equation}
\mathbf{e} = (e_x,e_y,e_z) = \left(\sin\vartheta\cdot\cos\varphi,\sin\vartheta\cdot\sin\varphi, \cos\vartheta \right)
\end{equation}
%maybe change to column vector
The resulting intensity for a transition between state $i$ and state $f$ excited by polarized light defined by the angles~$\vartheta$ and $\varphi$ would be:
\begin{equation}\label{eq:polarization-dependent-NEXAFS}
       \mu_{fi}^p(\vartheta,\varphi) = (D_{fi,x}\cdot e_x+D_{fi,y}\cdot e_y+D_{fi,z}\cdot e_z)^2
\end{equation}

For some systems, the variation of the azimuthal orientation $\varphi$ of the incident radiation relative to the substrate can also be used to extract structural information~\cite{Duncan2016}. Most systems, however, and this includes most metal-organic interfaces, possess a random azimuthal orientation, or many rotational domains which cancel out the $\varphi$-dependence due to the symmetry of the underlying substrate~\cite{Stoehr1992}. For such systems, it is therefore best practice to average over the azimuthal $\varphi$ dependence of the signal (in the $x,y$ plane) for the corresponding simulations, yielding \eref{eq:azimuthal-averaging-NEXAFS-1}:
\begin{equation}\label{eq:polar_ave}
 \mu_{fi}^{p,\mathrm{av}}(\vartheta) = \left(D_{fi,x}^2 + D_{fi,y}^2\right)\cdot\sin^2\vartheta  + D_{fi,z}^2\cdot \cos^2\vartheta
 \label{eq:azimuthal-averaging-NEXAFS-1}
\end{equation}

%%%%%%%%%%%%%%%%%%%%%%%%%%%%%%%%%%%%%%%%%%
\subsection{Computational Details}

The here presented DFT calculations have been performed with two different electronic structure software packages: the pseudopotential plane-wave code CASTEP~\cite{Clark2005} and the all-electron atomic-orbital code FHI-aims~\cite{Blum2009}. CASTEP was used to calculate NEXAFS spectra with the $\Delta$IP-TP method, as laid out in the method section and \fref{fig:workflow}. This approach includes $\Delta$SCF total energy calculations and TP calculations for each atom, all of which were performed in CASTEP. The all-electron code FHI-aims was used to calculate XPS energies with the $\Delta$SCF method to obtain comparable values free from the potential errors brought by the forced periodicity and the use of pseudopotentials in CASTEP. The raw inputs and outputs of all electronic structure calculations have been deposited in the NOMAD repository and are freely available (DOI:10.17172/NOMAD/2020.10.15-1).

\paragraph{All electron calculations in FHI-aims:} 
%%%%%%%%%%%%%%%%%HOW ARE DELTASCF calculations 
To localize the core-hole in the $\Delta$SCF calculation, several approaches were tested and are compared in \sref{section-core-hole}. These approaches employ the \emph{force\_occupation\_basis} (FOB) and the \emph{force\_occupation\_projector} (FOP) keywords implemented in FHI-aims by Matthias Gramzow (Fritz-Haber-Institute of the Max Planck Society, Berlin). Both keywords apply occupation constraints on KS eigenstates and use variants of the maximum overlap method (MOM) to ensure that the constraint is satisfied~\cite{Besley2009}. With the \emph{force\_occupation\_projector} keyword, we define the occupation constraint directly onto a KS eigenstate. If eigenstates swap between consecutive SCF steps, the constraint moves with the eigenstate following the maximum overlap with the constrained state of the previous SCF step. With the \emph{force\_occupation\_basis} keyword, we define the core-hole occupation constraint in terms of a localized atomic orbital basis function. The occupation constraint will be enforced on the KS eigenstate that has the highest contribution from the selected core atomic orbital. During successive SCF steps, the constraint will follow the eigenstate with the highest contribution from the atomic orbital, which ensures that the population of the core-level associated with the right atom is constrained. This approach works well for localized core-states such as 1s levels of most elements. The more delocalized the target core-state is, the less reliable this approach will be as basis functions on adjacent orbitals will complement each other. We find that the FOB approach is in most cases preferable as it correctly distinguishes between symmetry-equivalent atoms and therefore eliminates the necessity of an additional calculation to break the symmetry. We tested our approach against calculations with the method proposed by Kahk and Lischner~\cite{Kahk2019}, where the core-hole calculation using the \emph{force\_occupation\_projector}  functionality is enabled by the use of an initial wave function with reduced symmetry. This symmetry-broken wave function is created from a single SCF step with slightly increased core-charge on the relevant atom, resulting in a symmetry breaking for the desired core-state. This approach requires more effort then the FOB method, but proved more resilient when working with hybrid functionals. Alternatively, the issues of the FOB approach with hybrid functionals can be solved by combining it with a Boys localization that is implemented in FHI-aims~\cite{Rossi2016} and applied before the core-hole calculation~\cite{Foster1960,Michelitsch2019}.

The basis set convergence of the atom-centered numerical orbitals native to FHI-aims was extensively tested, as described in detail in the supplementary material (see tables S5-S8). To achieve a better description of the core-states, augmented near-core basis functions were introduced according to the literature~\cite{Kahk2019}. These additional basis functions are s-type hydrogen-like functions with high effective nuclear charges. Taking the results of the convergence series into account, the "tier2" basis with internal default tight settings for the numerical integration and the mentioned core-hole augmentation was chosen for the remaining calculations. The cluster calculations of the molecular crystal (in \sref{section-bulk-surface}) employed the "tight-tier2" basis set only for the atoms in the central molecule itself, while the surrounding 16 molecules were supplied with a "light-tier2" basis to reduce computational load. The same was true for the lower metal layers in the copper cluster calculations.

In \sref{section-xc}, different exchange correlation functionals were tested, these include PBE~\cite{Perdew1996}, PW91~\cite{Perdew1992b}, HSE06~\cite{Heyd2003}, PBE0~\cite{Adamo1999}, TPSS~\cite{Tao2003}, SCAN~\cite{Sun2015}, and xDH-PBE0~\cite{Zhang2012}.
The calculations were performed with the FHI-aims versions 200408 and 200511, both of which where checked to give the same results. 
All structures were optimized in FHI-aims with the PBE functional, a "tight-tier2" basis and a force threshold of 0.01~eV/\AA{}. The electronic convergence settings were set to $1\cdot 10^{-4}$~e/\AA{}$^3$ for the electron density, $1\cdot 10^{-2}$~eV for the KS-eigenvalues and $1\cdot 10^{-6}$~eV for the total energy. Relativistic effects were taken into account by employing the atomic ZORA functionality~\cite{Blum2009}. Higher order relativistic effects such as spin-orbit coupling could be neglected for the studied K-shell excitations and binding energies. XPS spectra were generated from the calculation results by representing each carbon 1s binding energy with a pseudo-Voigt function~\cite{Schmid2014,Schmid2014cor}. The application of this broadening scheme allows a better comparison to experimental data and enables a realistic judgement of the magnitude of differences in the XPS binding energies. The parameters of the pseudo-Voigt function were chosen to reflect typical experimental resolutions achieved in surface science experiments (FWHM = 0.7~eV, 70\%/30\% Gaussian/Lorentzian ratio).

\paragraph{Plane-wave calculations in CASTEP:}
The plane-wave code CASTEP was used to perform both $\Delta$SCF and TP calculations within the $\Delta$IP-TP approach. All calculations were carried out in periodic boundary conditions (PBC) using on-the-fly generated pseudopotentials. The stability of the calculation results with regards to the additional parameters introduced by PBC and pseudopotentials was extensively tested. In particular the choice of pseudopotentials, the energy cut-off for the pseudopotentials, and the vacuum box size around isolated molecules (necessary due to the PBC), was investigated in detail, as presented in \sref{section-basis}. Calculations were performed using either CASTEP version 18.1 or 19.11~\cite{Clark2005}, while maintaining the same pseudopotential definitions and with a total energy per atom convergence tolerance of at least $1\cdot 10^{-6}$~eV/atom. For the calculation of XPS binding energies, the $\Delta$SCF method was used. Here, for each carbon atom an excited core-hole calculation was performed, where the electron configuration of the respective pseudopotential was modified  to  [1s$^1$, 2s$^2$, 2p$^3$], effectively localising the core-hole at this atom while moving the electron to the valence region. However, the new electron in the valence region is then removed by the introduction of a total charge of +1.0~e. Due to the periodic boundary conditions, this charge is compensated by a homogeneous background charge of -1.0~e, yielding a net-neutral unit cell. The core-level binding energies obtained for each carbon atom have furthermore to be corrected for the energy change of the core-electrons contained in the modified pseudopotentital~\cite{Mizoguchi2009}. NEXAFS simulations were carried out using the ELNES module in CASTEP~\cite{Mizoguchi2009}, the half core-hole was included by changing the electron configuration of the pseudopotential to [1s$^{1.5}$, 2s$^2$, 2p$^{2.5}$]. Each (half) core-hole calculation includes a total energy SCF calculation followed by a band structure calculation to converge the unoccupied states, and the subsequent evaluation of the dipole matrix elements using the projector augmented wave reconstruction method to reconstruct  the core-electron wave functions~\cite{Gao2009}. A self-written post processing tool called MolPDOS~\cite{Maurer2013} was used to postprocess the data and generate the spectra. This tool is contained in CASTEP version 18.1 and later and exists alongside other useful tools for this purpose such as OptaDOS~\cite{Nicholls2012}. The total NEXAFS spectrum is produced from summing all single carbon species contributions and shifting them with their corresponding core binding energies, resulting in the so-called $\Delta$IP-TP approach.

To make the NEXAFS spectra obtained by the calculations comparable to experimental data, an photon energy dependent broadening scheme using pseudo-Voigt functions~\cite{Schmid2014,Schmid2014cor} was included. As laid out in \sref{methods-section}, the change in peak shape and width is meant to simulate the different life-times of excited states below and above the ionization potential. After the correction according to the XPS energies, the NEXAFS spectrum is divided into three ranges. The first range starts with the leading edge, spans the first 5~eV of the spectrum and is assigned a pseudo-Voigt FWHM of 0.75~eV and a 80\%/20\% Gaussian/Lorentzian ratio, while the third range starts from 15~eV above the leading edge and is assigned a FWHM of 2.0~eV and a 20\%/80\% G/L ratio. Both ranges are connected by an intermediate range, in which the FWHM and the G/L ratio change linearly. Every NEXAFS transition is represented by a pseudo-Voigt peak with parameters according to its position on the photon energy axis. The values mentioned here were consistently used in this publication. If comparison to experiment is desired, these broadening parameters need to be adjusted accordingly, e.g. as done in figure S6 of the supplementary material. 

\section{Results and Discussion}

In the following, we will discuss the calculation of X-ray absorption and photoelectron spectra for a small set of typical systems to exemplify the technical and numerical aspects, as well as the pitfalls involved in performing such calculations. We performed calculations for azupyrene (see \fref{fig:workflow}b) and ethyl trifluoroacetate (ETFA, see \fref{fig:xc_XPS}a), as well as for the gas-phase azulene molecule, the azulene bulk crystal, and azulene adsorbed on a Cu(111) surface (see \fref{fig:azulene_system_comparison})~\cite{Klein2019}. ETFA shows extreme chemical shifts in its carbon 1s binding energies and has been an important reference system since the dawn of photoelectron spectroscopy~\cite{Siegbahn1973,Gelius1973,Travnikova2012}. Azupyrene is a polycyclic aromatic hydrocarbon, which was chosen because of its high symmetry, yielding a large number of equivalent carbon atoms. The three different systems involving the azulene molecule are used to show the versatility of our approach. A comprehensive experimental data set exists for azulene on Cu(111) which was previously used to show the feasibility of our DFT-based structures, energies and spectroscopic simulations~\cite{Klein2019,Klein2019a,Kachel2020}.

\subsection{\label{section-core-hole} The choice of core-hole localization method for XPS simulations}

%start with introducing the model systems that we will look at.
%then explain which atoms are symmetry equivalent

\begin{figure}
    \centering
    \includegraphics[width=0.7\linewidth]{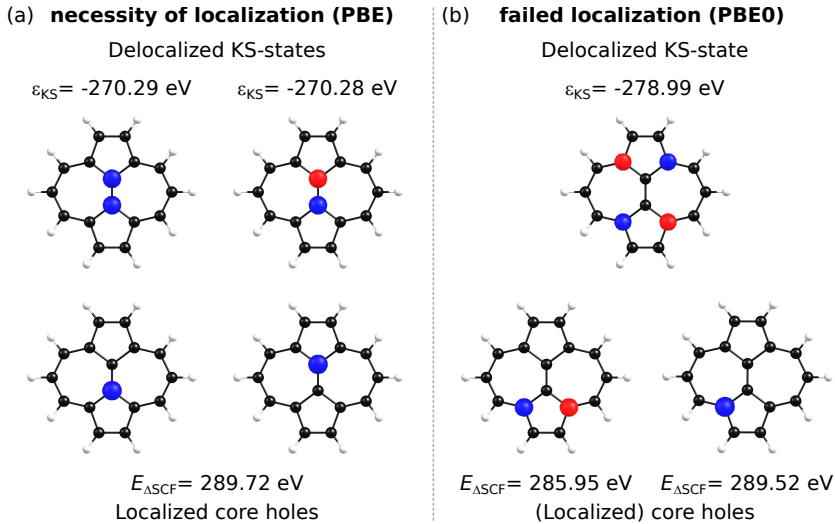}
    \caption{Core-hole localization for azupyrene. (a) The KS states of the two central C 1s orbitals are delocalized over both atoms, resulting in two KS energies, which cannot be assigned to one carbon atom (upper part). The correctly localized core-holes in the $\Delta$SCF calculation yield the same binding energy for both symmetry equivalent carbons (lower part). (b) The core-hole localization, if not strictly enforced, often fails for hybrid functionals (e.g. PBE0). In this example, the localization worked for one case and produced a core-state properly localized on one carbon atom with a realistic binding energy (289.52 eV), while for the second case the core-state is still delocalized over two of the four symmetry equivalent atoms and the binding energy is unrealistically low (285.95 eV). The iso-value for all states shown is 0.1.}
    \label{fig:localization}
\end{figure}

The most straightforward way to calculate XPS energies via DFT is the use of the negative KS orbital energy values $\varepsilon_\mathrm{KS}$ according to \eref{eq:koopmans}. In addition to neglecting  relaxation effects, this approach carries an additional problem: The KS-states may be delocalized, especially for highly symmetric molecules, e.g.  azupyrene, which possesses the symmetry point group $D_{2h}$. In such cases, the KS states for symmetrically equivalent atoms are linear combinations of the atomic core-states (see~\fref{fig:localization}). If the respective atoms are close enough to produce appreciable overlap, an issue is encountered: The energy of the bonding combination of the core-states is lowered while the energy of the anti-bonding combination is raised. Therefore, two different binding energies exist for two symmetrically equivalent carbon atoms (see \fref{fig:localization}a). 

The delocalization of the KS-states is not only an issue for the direct use of the KS  energies to model binding energies, but also in $\Delta$SCF calculations that account for core-hole screening effects, \eref{eq:deltascf}. If the occupation of a certain core-level is changed to calculate the ionized state for the $\Delta$SCF calculation, for example by using the {\it{force\_occupation\_projector}} functionality in FHI-aims to remove an electron from a KS state, it might distribute the core-hole over multiple atoms. The result is an unphysical system with a calculated binding energy between the values of KS-states and the properly localized core-holes (see \fref{fig:localization}b).

To avoid these issues, additional steps have to be taken to assure the proper localization of the core-hole in the calculations of the ionized state for the $\Delta$SCF approach. In particular, it is necessary to break the symmetry of the electronic wave function of the system. In a pseudopotential code this step is straightforward, because the use of a pseudopotential with a built-in core-hole for each atom automatically breaks the symmetry and localizes the core-hole~\cite{Gao2009}. The same is true for frozen core calculations, where the ionic occupation for the core-orbitals is frozen out with a core-hole localized at the desired atom~\cite{Chong2002}.

For all-electron calculations, the problem is more difficult to solve. One approach suggested by Kahk and Lischner~\cite{Kahk2018,Kahk2019} is to introduce a third step in addition to the ground state and the core-hole calculation. In this intermediate step, the symmetry of the wave function is broken, e.g. by introducing an increased core-charge to the atom in question. The (unphysical) wave function with  broken symmetry and increased core-charge is then used as starting point for a core-hole calculation  with normal core-charge at all atoms~\cite{Kahk2018,Kahk2019}. The \emph{force\_occupation\_projector} functionality in FHI-aims is able to implement this approach. 
Alternatively, localization can also be achieved by using the \emph{force\_occupation\_basis} functionality in FHI-aims. In the latter, the occupation constraint on the KS state is defined by its overlap with a 1s atomic orbital basis function localized at the relevant atom. This approach in most cases automatically breaks the symmetry of the wave function during the self-consistent core-hole calculation.

It should be noted that both approaches, if the core-hole localization is successful, yield numerically identical results for the systems studied here. For most cases, we have found the \emph{force\_occupation\_basis} approach to be more stable with regard to electronic convergence and more convenient, because it does not require an additional initialization step. However, in the case of the highly symmetric azupyrene molecule, the direct use of the \emph{force\_occupation\_basis} approach failed  for all hybrid functionals (see \fref{fig:localization}b). To obtain the correct binding energies when using hybrid functionals it was necessary to first break symmetry using the Boys localization~\cite{Foster1960}, as proposed in the literature~\cite{Michelitsch2019}. This approach was successful for all molecule and XC functional combinations.
It appears that hybrid functionals, contrary to GGA and meta-GGA functionals, suffer much more from variational collapse during $\Delta$SCF calculations and require a constraint that explicitly breaks symmetry to ensure core-hole localization. An additional way for solving this problem would be the use of a modified version of the $\Delta$SCF method, such as the local SCF method~\cite{Ferre2002,Loos2007} or the linear-expansion $\Delta$SCF~\cite{Maurer2013} approach. 
Additionally, one of the tested meta-GGA functionals (SCAN) showed unphysical behaviour with great differences in the valence spin channels caused by the core-hole. This behaviour occurred both when using \emph{force\_occupation\_basis} and \emph{force\_occupation\_projector} and the reason for its occurrence is still unclear. 
For the unsymmetrical molecule ETFA  (symmetry point group $C_1$ in the chosen rotational conformation), the localization poses less of a problem. Here, both tested meta-GGA functionals (SCAN and TPSS) failed when using the \emph{force\_occupation\_basis} approach for some atoms, but results could be obtained using the \emph{force\_occupation\_projector}.

%%%Head-Gordon paper~\cite{Hait2020}

The localization of the core-hole is also important for the calculation of NEXAFS spectra with all-electron DFT and was recently discussed in detail in the literature~\cite{Michelitsch2019}. In this work, we only perform NEXAFS simulations in the plane-wave code CASTEP with core-hole excited pseudopotentials. Here, the description of the (half) core-hole within the pseudopotential automatically ensures localization and symmetry breaking. The additional effects this core-hole excited pseudopotential exerts on XPS and NEXAFS signatures is discussed separately below.

\subsection{\label{section-xc} The choice of the exchange correlation functional}

\begin{figure}
    \centering
    \includegraphics[width=\linewidth]{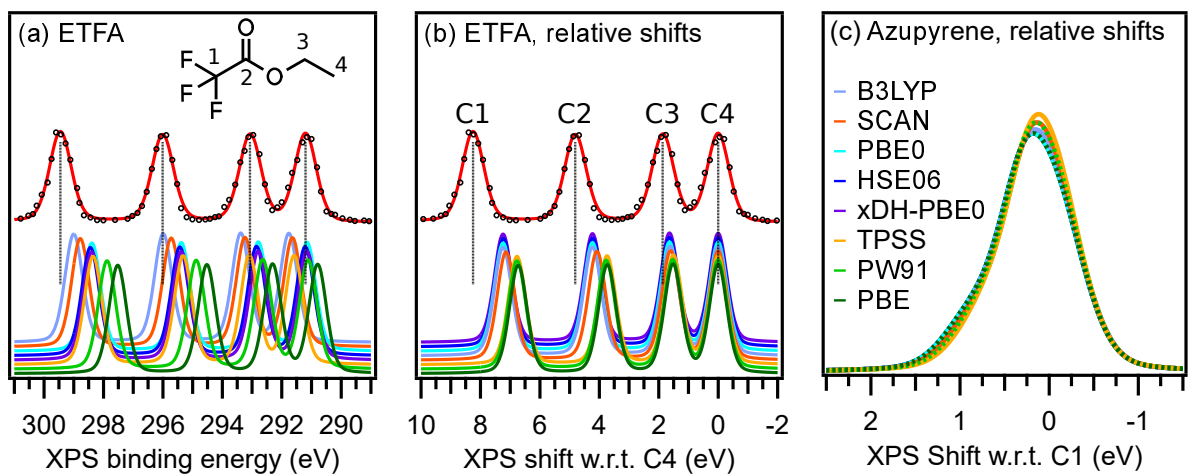}
    \caption{Comparison of the C 1s XPS spectra of ETFA and azupyrene  calculated with different XC functionals. (a) XPS spectrum of ETFA on the absolute binding energy scale with respect to the vacuum level. Upper part: experimental spectrum from reference~\cite{Gelius1973} (open circles, data points; red line, fit function). Lower part: calculated spectra. Vertical lines as guide-for-the-eye. (b) XPS spectrum of ETFA on a relative binding energy scale w.r.t. the atom C4.  (c) XPS spectrum of azupyrene on a relative binding energy scale w.r.t. C1 (see \fref{fig:workflow}b).  
    Both spectra were obtained by applying the same pseudo-Voigt broadening on the calculated C 1s binding energies to mimic a typical experimental resolution. Inset: structural formula of ETFA with numbering. The color scheme for the various functionals in (c) is valid for all panels.}
    \label{fig:xc_XPS}
\end{figure}

To judge the influence of the exchange correlation functional on the absolute and relative C 1s XPS energies, a comprehensive study sampling 8 functionals of various types was carried out. The tested functionals included GGAs (PBE and PW91), meta-GGAs (SCAN and TPSS), hybrid functionals (PBE0, HSE06 and B3LYP), as well as the double hybrid functional xDH-PBE0. The test calculations were performed on the two gas phase molecules ETFA and azupyrene. These two molecules where chosen because they represent two different challenges when calculating XPS binding energies. ETFA (see \fref{fig:xc_XPS}a) is commonly used as example for XPS measurements and calculations due to the extreme chemical shifts of its four different carbon atoms~\cite{Siegbahn1973,Gelius1973,Travnikova2012}. On the other hand, azupyrene (see \fref{fig:workflow}b) represents an alternative challenge in XPS calculations. It is a common organic molecule with carbon atoms in similar environments that produce slight but possibly measurable relative shifts in their C 1s binding energies. For both molecules, the calculations regarding the performance of the XC functionals were carried out using the all-electron code FHI-aims with a "tight-tier2" core-augmented basis set.

The ETFA molecule possesses extreme chemical shifts for the C1s binding energies of up to 8 eV, leading to clearly separated peaks in a spectrum with typical experimental resolution. There is high quality experimental reference data available obtained by measurements on the gas-phase ETFA molecule~\cite{Gelius1973,Travnikova2012}. This experimental data is compared to the spectra calculated with different XC functionals in \fref{fig:xc_XPS}a on an absolute binding energy axis. For reference, all calculated binding energies are also compiled in table S1 of the supplementary material.

On first sight, the overall performance in terms of absolute binding energies is not straightforward to extract (\fref{fig:xc_XPS}a, data tabulated in table~S1). However, a clearer trend emerges when the spectra are plotted as relative shifts with respect to the lowest binding energy (\fref{fig:xc_XPS}b, data tabulated in table~S2).
The performance of the different functionals regarding the relative shifts in the binding energies can be divided into three groups: (1) PBE, PW91, TPSS with average deviations of 0.74, 0.72, 0.73~eV; (2) SCAN with 0.53~eV; (3) PBE0, HSE06, B3LYP, xDH-PBE0 with 0.45, 0.46, 0.46 0.45~eV, respectively. It is apparent that the hybrid functionals perform best here, and the GGAs perform worst, while TPSS behaves more like the GGAs and SCAN is somewhere in between.

When discussing the performance on the absolute binding energy scale, things are more complex. Due to an additional global displacement of all calculated energies, the average deviation in the absolute binding energies would seem to be smallest for B3LYP and SCAN. But according to the relative shifts, the case can be made that all functionals have similar problems describing the binding energies of the carbon atoms with extreme binding energy shifts. In fact, the functionals PW91, PBE0, HSE06 and xD-PBE0 have an excellent agreement with the absolute binding energy of carbon C4, which possesses the lowest binding energy. Following this argument, the best performance for the absolute binding energies is provided by the hybrid functionals PBE0, HSE06 and xD-PBE0 with an average deviation of 0.50, 0.46, 0.50~eV and the GGA PW91 with the deviation of 0.81~eV. 

By comparing absolute and relative deviations, it is obvious that the main error in these calculations is not only in the absolute energies, but already in the relative energies. Generally the hybrid functionals perform slightly better than the meta-GGAs and GGAs but their results are still quite poor compared to experimental data. Apparently all functionals have problems calculating the high binding energies resulting if the core-state is extremely descreened, e.g. for the carbon atom C1 in the CF$_3$ group of ETFA. For carbon atoms in more common chemical environments, e.g. C4 in ETFA, all functionals perform satisfactorily, i.e. yielding a deviation below or around 0.5~eV. It should also be noted that the use of the double hybrid functional xD-PBE0 brings no improvement relative to the regular hybrids. The better performance of hybrid functionals for K-shell XPS energies was also reported in the literature~\cite{Pi2020}, while other studies found a better performance of regular GGAs~\cite{Tolbatov2016}. However, many test sets to judge the performance of binding energy calculations include only molecules with more moderate shifts ~\cite{Chong2007,Triguero1998,Tolbatov2016,Pi2020}.

\begin{table*}[]
\caption{Relative shifts of the carbon 1s binding energies of the azupyrene molecule in dependence of the XC functional, calculated using the $\Delta$SCF approach in the all-electron code FHI-aims. All shifts are in eV and w.r.t. the binding energy of C1 (see \fref{fig:workflow}b).}
\footnotesize
\lineup
\begin{indented}
\item[]\begin{tabular}{lcccccccc}
\br
Carbon & PBE & PW91 & HSE06 & PBE0 & B3LYP & SCAN & TPSS & XDH-PBE0 \\\mr
C1     & 0.00     & 0.00      & 0.00       & 0.00      & 0.00       & 0.00      & 0.00      & 0.00          \\
C2     & 0.25     & 0.25      & 0.31       & 0.32      & 0.31       & 0.26      & 0.25      & 0.32          \\
C3     & 0.25     & 0.25      & 0.26       & 0.26      & 0.22       & 0.23      & 0.19      & 0.26          \\
C4     & \-0.18    & \-0.19     & \-0.19      & \-0.19     & \-0.20      & \-0.19     & \-0.18     & \-0.19         \\
C5     & 0.82     & 0.82      & 0.90       & 0.91      & 0.85       & 0.86      & 0.77      & 0.91   \\\br
\end{tabular}
\label{tab:azupyrene_rel}
\end{indented}
\end{table*}

To further test the performance of the XC functionals for less extreme chemical shifts, we turn to the molecule azupyrene, which is an aromatic hydrocarbon without hetero-atoms. The chemical shifts exhibited by its carbon atoms are therefore more subtle and only caused by its non-alternant topology, similar to the azulene molecule already discussed in the literature~\cite{Klein2019}. To compare the performance of the functionals with regard to these more subtle shifts, only relative differences in the binding energies are analysed. These relative shifts with respect to carbon C1 (see \fref{fig:workflow}) are compared in \tref{tab:azupyrene_rel} for the different XC functionals. For reference, the absolute binding energies can be found in table S3.
Within each class of functional used, GGA, meta-GGA and hybrid, the values obtained for the relative XPS shifts are similar with maximal deviations of 0.06 eV within the hybrids and 0.09 eV in the meta-GGAs. As a caveat, the meta-GGA SCAN showed problems with the response of the valence electron states to the core-holes and produced an unphysical spin density. However, the C1s shifts are still in line with the other functionals. It should be noted that the simple approach using Koopmans' theorem also gives similar relative shifts in the binding energies (see table S4 in the supplementary material.) This indicates that for the azupyrene molecule the shifts are mainly determined by the chemical environment of the initial state and that the relaxation of the core-hole is very similar for all chemical species.

\Fref{fig:xc_XPS}b shows the XPS spectrum of azupyrene obtained by applying the same pseudo-Voigt broadening as used in \fref{fig:xc_XPS}a to all relative binding energies in \tref{tab:azupyrene_rel}. When the results obtained with the different functionals are thereby compared against the backdrop of a typical experimental resolution, they provide almost indistinguishable results. For this reason, we chose the functional PBE for all subsequent calculations.

\subsection{\label{section-basis} The choice of basis set in XPS and NEXAFS simulations}

In sections \ref{section-core-hole} and \ref{section-xc} we have discussed the role of core-hole localization and exchange-correlation functional when performing all-electron atomic orbital-based core-hole simulations of XPS binding energies. Another important numerical choice is the type and size of basis set that is used for such calculations. For atomic orbital basis sets, the importance of core augmentation functions has been extensively discussed in literature~\cite{Besley2009,Kahk2019,Ambroise2019,Sarangi2020}. We also find that basis set convergence for XPS binding energies in FHI-aims is very fast and, once core augmentation functions (according to reference~\cite{Kahk2019}) are added, does not require to go beyond the standard 'tight' basis set definitions. The full data for the convergence series is shown in tables S5 to S8 in the supplementary material.

Moving from an all-electron basis to a pseudopotential plane wave formalism has the benefit of eliminating the issue of core-hole localization that was discussed in \sref{section-core-hole}, because core-hole excited pseudopotentials account for the core-excitation only on the relevant atom in question. However, this introduces other challenges. For example, the frozen core described by the pseudopotential does not relax and therefore absolute binding energies are typically much worse than if all-electron descriptions are used. Similarly, as the core-states are not explicitly treated, the core-level wave function needs to be rebuilt to enable the calculation of transition dipole moments between core and valence states in NEXAFS~\cite{Gao2009}. Lastly, it is not clear how the choice of pseudopotential affects core-level binding energies and XAS signatures.

Different types of pseudopotentials exist varying in their degree of  "hardness", i.e. the number of states that are pseudoized and the constraints that are placed on the core region~\cite{Payne1992}. Examples include non-local norm-conserving pseudopotentials~\cite{Kleinman1982} and ultra-soft pseudopotentials~\cite{Vanderbilt1990}. The latter require much lower energy cutoffs and fewer plane waves to accurately represent observables, which they achieve by lifting the constraint of charge conservation in the core region and making wave functions non-orthogonal. 
We compared XPS and NEXAFS spectra calculated with three different pseudopotentials for the example molecule azupyrene. Those three are based on the default settings of on-the-fly generated pseudopotentials in CASTEP and only differ in the type of projector, namely they use (for each angular momentum channel): (1) a norm-conserving projector, (2) a single ultrasoft projector, or (3) two ultrasoft projectors. All other parameters that define the pseudopotentials were left unchanged from the defaults. The detailed definitions are available in the raw data deposited on the NOMAD repository.

For the XPS calculations, we compared the spectra obtained with all three pseudopotential definitions with the spectrum generated by the all electron code FHI-aims. The XPS spectra of the different pseudopotentials had to be rigidly shifted by -6.1 eV to align them to the all-electron results, because the lack of electron relaxation imposed by the frozen core approximation and the introduced charge background discussed below. The peak shape, however, is almost identical for all types of projectors and the all-electron method. The visualization of this result is shown in figure~S1a and the relative peak positions of the different carbon species are compiled in table~S9 of the supplementary material. In the case of NEXAFS simulations, the same is true and the different pseudopotentials provide almost identical spectra, with the exception of slight differences in intensities at high excitation energies (see figure~S1b of the supplementary material). We therefore find that the main features of core-level spectra are robust with respect to the type of pseudopotential projection functions used. For all further calculations, we select the ultrasoft pseudopotential with two projectors per angular momentum channel, which is the default in CASTEP.

Another consequence of using a plane-wave pseudopotential basis set for core-hole simulations is the limitation that all calculations have to be performed using periodic boundary conditions (PBCs). The use of PBCs with a final state core-hole simulation approach such as $\Delta$SCF or TP can introduce errors in the calculation of XPS binding energies. Errors can arise due to finite size effects, when unit cells are too small and core-holes are interacting across periodic images. Another issue is when the positive charge of the core-hole is not compensated. $\Delta$SCF simulations of XPS binding energies and $\Delta$IP-TP simulations of NEXAFS transitions introduce a net charge of +1 or +0.5, respectively. Unit cells in PBCs must be charge neutral, otherwise the electrostatic potential would diverge. Therefore, a net charge is typically compensated by the introduction of a uniform (negative) background charge over the whole unit cell. This is a fail safe mechanism of an Ewald summation rather than an artefact, although it does introduce erroneous physics, such as a strong dependence of the total energy on the size of the unit cell.

\begin{figure}
    \centering
    \includegraphics[width=0.7\linewidth]{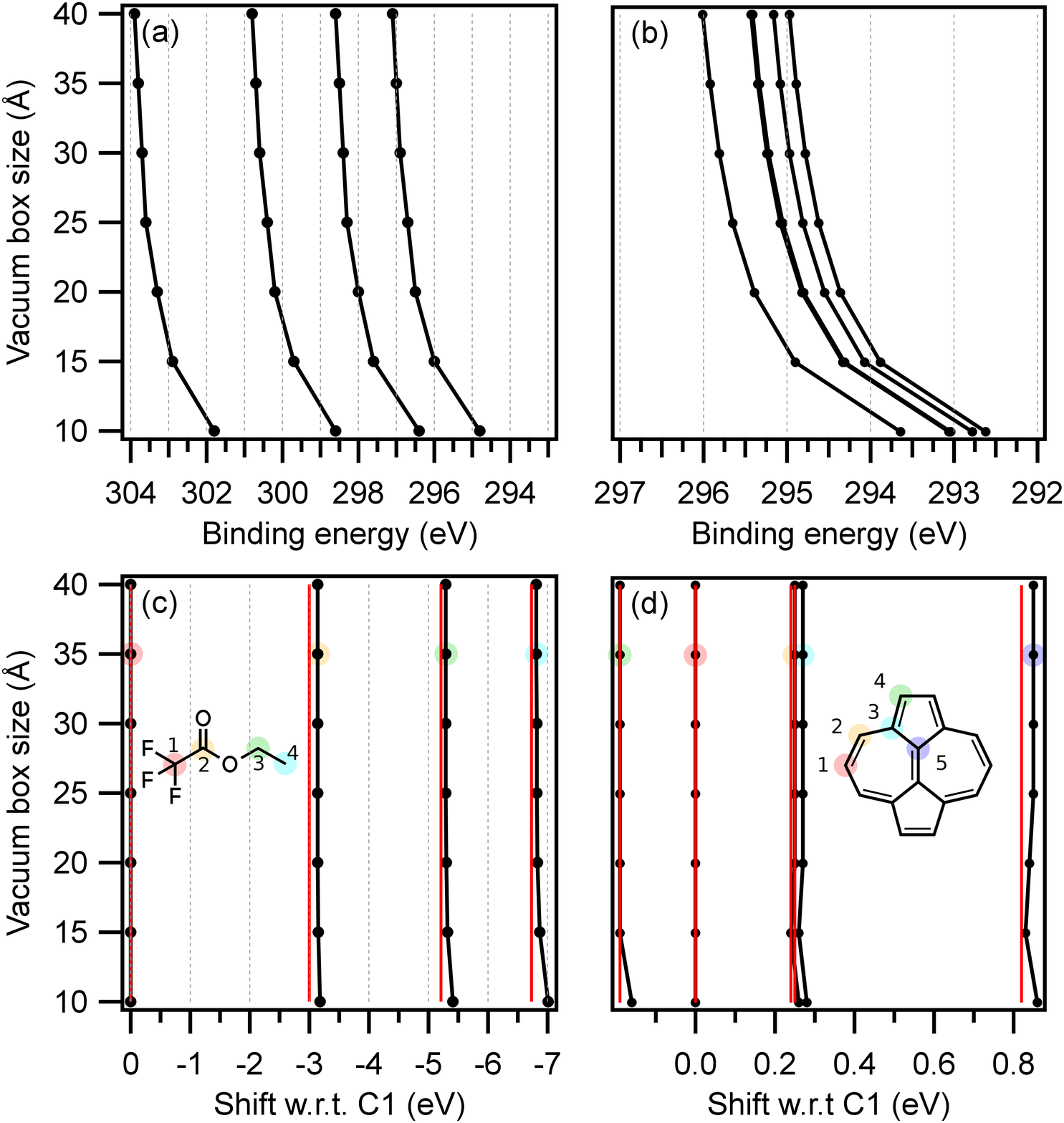}
    \caption{Graphs comparing the influence of the vacuum box size on the XPS binding energies of gas-phase molecules in periodic calculations. ETFA on the left and azupyrene on right. (a) and (b) show the changes in the absolute binding energy of C 1s electrons. (c) and (d) show the relative shifts in C1s binding energy with respect to C1. Lines in red represent the relative shifts obtained from the all-electron FHI-aims calculations without periodic boundary conditions and the grey gridlines highlight 1 eV energy steps. All calculations were performed with the PBE functional.}
    \label{fig:XPS_box_size}
\end{figure}

To calculate gas-phase molecules with PBC, the molecule is placed in a large vacuum box (a so-called supercell). \Fref{fig:XPS_box_size} shows the convergence behaviour of the C 1s binding energies for the gas-phase ETFA and azupyrene molecules as a function of the size of this vacuum box. The box is of cubic shape with increasing size from 10 to 40 \AA{} in all directions. It can be seen that the absolute values of core-level binding energies (\fref{fig:XPS_box_size}a,b) vary by 2-3 eV as the box size increases and are clearly not converged even at very large vacuum box sizes. This effect can be attributed to the change in core-hole density and electrostatic interaction with the background charge and was recently studied in detail by Taucher \emph{et al.}~\cite{Taucher2020}. However, this dependence on the box size of the absolute energies is not seen in the relative chemical shifts between the single carbon atoms within the molecules. In \fref{fig:XPS_box_size}c,d the behaviour of the relative shifts with the vacuum box size is plotted and compared to the relative shifts obtained from the all-electron code FHI-aims (in red). Here it is clear that the relative shifts are much less affected by the background charge. While very small box sizes lead to slight deviations in relative shifts, the peaks converge very quickly and after about 20 \AA{} box size, no change is seen. The converged relative shifts agree well with the all-electron calculations with average deviations of 0.1 eV for the extreme shifts in ETFA and 0.01 eV for the more subtle shifts in azupyrene. This result shows that absolute binding energies are affected by a homogeneous background compensation charge, but the issue is much less severe for relative energies where careful convergence can reproduce all-electron results. In addition to this data, we compare the performance of the $\Delta$SCF and $\Delta$IP-TP methods with charge neutral variants for azulene adsorbed on Cu(111) in \sref{section-bulk-surface}.

The influence of the box size was also studied for the NEXAFS spectra calculated with the $\Delta$IP-TP method and shown in figure~S2a of the supplementary material. As seen with the XPS binding energies, the NEXAFS spectra are converged at a box size of 20~\AA{} where little to no change is seen afterwards. Regarding basis set convergence, the spectra show little change in the general shape and features with increased cut-off energy (see figure S2b), while the overall intensity increases and is converged after 450 eV, which is a plane wave cut-off energy similar to what is typically required for intermolecular interaction energies or adsorption energies with the employed ultrasoft pseudopotentials. We complete our discussion of numerical and technical parameters of core-hole simulations by studying the influence of the exchange-correlation functional on the simulated NEXAFS spectrum. We have calculated the NEXAFS spectrum of azupyrene with three different GGA functionals: PBE, BLYP and PW91 (see figure~S3  of the supplemental materials). While there are slight differences for the intensities of higher energy transitions, the spectra show little variation overall and are almost identical for the lowest energy peaks. The deviations found by us are slightly smaller than what has previously been reported~\cite{Triguero1998}.

\subsection{\label{section-atom-MO} Analysis of core-level spectra in terms of initial and final state contributions}

\begin{figure}
    \centering
    \includegraphics[width=0.3\linewidth]{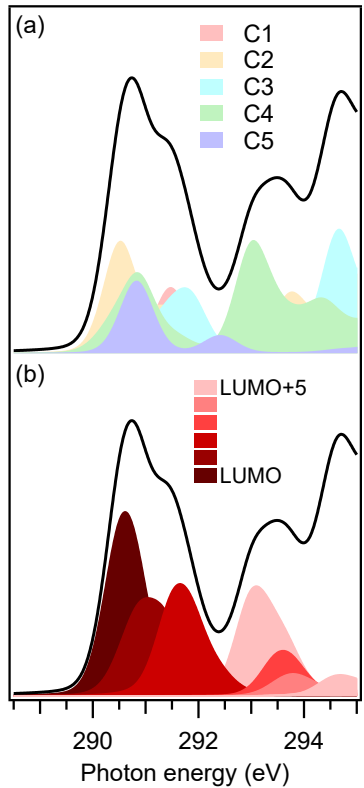}
    \caption{Different decomposition schemes for the simulated NEXAFS spectrum of azupyrene. (a) The initial state decomposition reveals which part of the spectrum is related to which carbon 1s orbital. For the presented case of the highly symmetric azupyrene molecule, the groups of symmetry equivalent atoms were summed up to improve visibility. (b) The final state decomposition shows the contributions of the transitions to each unoccupied molecular orbital.}
    \label{fig:final-initial}
\end{figure}

Once XPS and XAS spectra are simulated, a comparison with experiment often requires a deeper analysis of the individual signatures that constitute the spectra. The decomposition in terms of the core-levels from which photoelectrons and excited electrons originate is trivial, as our approach uses separate core-hole calculations for each carbon atom (see \fref{fig:workflow}). Therefore initial state decomposition is achieved by only summing up the transitions which originate from the 1s orbital localized at a specific carbon atom as shown in \fref{fig:final-initial}a for the azupyrene molecule.

An analysis in terms of final states in NEXAFS simulations requires further consideration, but can be particularly appealing for organic molecules adsorbed at surfaces and in thin-films, as spectra are often interpreted in terms of the symmetry of these states ($\pi^*$, $\sigma^*$). The final unoccupied state decomposition for azupyrene in \fref{fig:final-initial}b shows the spectrum projected onto molecular orbital ($\mathrm{MO}$) reference states according to their overlap with the final states $f$ of each transition~\cite{Maurer2013}. 
 \begin{equation}\label{eq:c_MO}
    c^\mathrm{MO}_{f} =  \left|\langle\psi^\mathrm{MO}|\psi_f \rangle \right|^2
 \end{equation}
The projection coefficient $c^\mathrm{MO}_{f}$ is then multiplied with the cross section in \eref{eq:NEXAFS2} to select core-level transitions that lead to an excitation of a final state that has a contribution from the reference state $\mathrm{MO}$:
\begin{equation}\label{eq:I_MO}
     \sigma_i^{MO}(h\nu) \propto  \sum_f  c^\mathrm{MO}_{f} \left|\mu_{fi}\right|^2 \delta(h\nu-\Delta E_{fi})
 \end{equation}
This approach is implemented in CASTEP via the MolPDOS tool~\cite{Maurer2013}, which enables such projections onto densities-of-states. For the projection, the reference states $\mathrm{MO}$ have to be provided for each TP calculation in the form of a reference wave function. In the simplest case, this reference wave function is generated by the ground state calculation of exactly the same system as used for the transition potential calculations (as shown in \fref{fig:final-initial}b). In this case the projection scheme directly provides contributions of each unoccupied state to the overall NEXAFS spectrum.

However, it might be beneficial to choose a different reference system. For example, if the overall system in question is a molecule adsorbed on a metal surface, the reference system can be chosen as the free standing molecular overlayer (in exactly the same structure but without the surface). The projection of the NEXAFS transitions of the whole adsorbate system on the unoccupied states of the molecular overlayer now yields information about how the molecular states are changed by the adsorption and how they still contribute in a distinct manner to the overall spectrum. This approach was already used multiple times and proved its utility for molecules adsorbed on metal surfaces both with regard to NEXAFS transitions and for the analysis of the ground-state density of states~\cite{Mueller2016,Diller2017,Klein2019,Klein2019a,Klein2020}.

\subsection{\label{section-bulk-surface} XPS and NEXAFS simulation of organic crystals and metal-organic interfaces}

\begin{figure}
    \centering
    \includegraphics[width=0.8\linewidth]{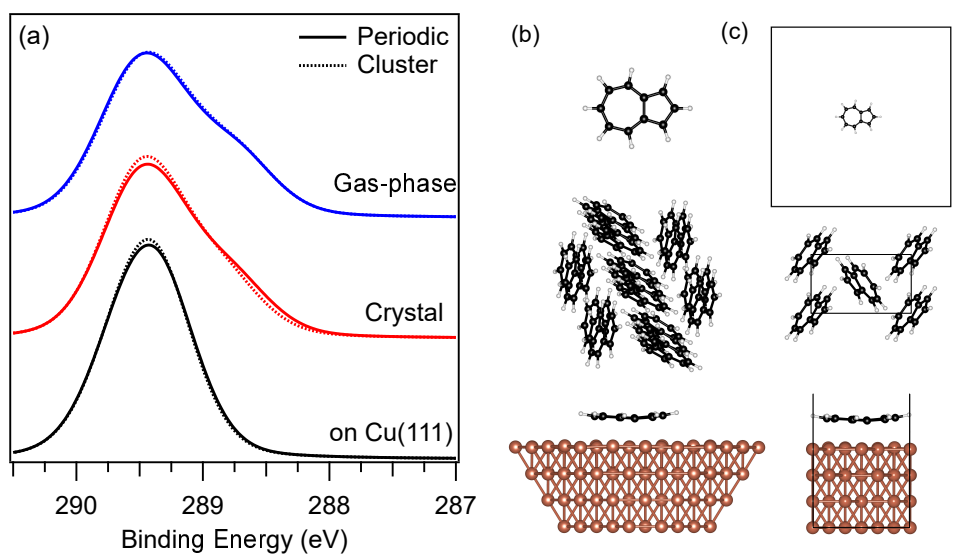}
    \caption{XPS calculations of azulene, as a free molecule, in the molecular crystal, and when adsorbed on Cu(111). (a) Comparison of the XPS spectra, calculated both with the cluster approach in the all-electron code FHI-aims and in a periodic system with the plane-wave code CASTEP. The XPS spectra were shifted to align the peak maxima and to better show the difference in the peak shape; broadening was introduced to simulate a typical experimental resolution. Right side: Structures of the used clusters (b) and periodic unit cells (c).}
    \label{fig:azulene_system_comparison}
\end{figure}

Up to now, we have only discussed core-level simulations of XPS and NEXAFS spectra for isolated molecules. In the following, the three different systems involving the azulene molecule are used to show the versatility of our approach and pinpoint special challenges for spectroscopic calculations of bulk and surface systems. As exemplar systems, we chose the azulene molecule both in a molecular crystal and adsorbed on the Cu(111) surface, depicted in \fref{fig:azulene_system_comparison}, and compare both to the gas-phase molecule. 

One additional issue of condensed systems compared to gas-phase calculations is the sampling of electronic states across the periodic crystal via Brillouin zone integration (or k-space sampling). Therefore, the convergence of both the XPS and NEXAFS calculations has to be ensured with respect to the number of k-points included. In the supplementary material (see figures S4 and S5a), we provide a detailed discussion of the convergence properties of the XPS and NEXAFS spectra for the azulene molecular crystal and azulene adsorbed on Cu(111) when simulated with CASTEP. In short, both systems show rapid convergence with respect to the density of the employed Monkhorst-Pack k-grid~\cite{Monkhorst1976}. In the case of azulene on Cu(111), we have additionally tested convergence with respect to the size of the vacuum slab that separates the surface model from its periodic image perpendicular to the surface and find that this parameter shows virtually no influence on the spectra (see figure S5c). Moreover, we also repeated the convergence series with regard to the  cutoff energy of the plane wave basis and found that the convergence of the metal-adsorbed molecule is similar to the gas-phase molecule (see figure S5b). 

The comparison between gas-phase azulene, the azulene molecular crystal and azulene adsorbed on Cu(111) allows us to show that all three systems can be treated with approximate core-hole simulation methods such as $\Delta$SCF and $\Delta$IP-TP, but also allows us to compare aperiodic simulations with FHI-aims and calculations under PBCs with CASTEP. First, XP spectra where calculated with the $\Delta$SCF approach in both FHI-aims and CASTEP. In the case of FHI-aims, we employ an aperiodic approach (also called cluster approach) where we correctly account for the positive charge in all three systems after photoionization. On the other hand, the plane wave nature of CASTEP made a 3D periodic cell necessary, even for those systems showing a lower periodicity. The cluster approach in FHI-aims can easily simulate the isolated molecule, while the inherently 3D periodic plane-wave approach has to simulate the molecule in a large periodic vacuum box (see \sref{section-basis} for a discussion) at significant computational overhead. For the molecular crystal and the molecule adsorbed on the surface, the plane wave code is more suitable, because it already includes the 3D/2D periodicity. In the cluster approach for the all electron calculation of the crystal and the metal-organic interface, the choice of the cluster is not straightforward. For the molecular crystal, the 16 molecules surrounding a central molecule were cut out of the periodic crystal structure and the XPS shifts of the central molecule were calculated. For the surface-adsorbed molecule, a cluster was cut-out of the 2D periodic slab. The choice of the cluster was guided by the desire to preserve the hexagonal symmetry of the surface while simultaneously providing a sufficient surface area for adsorption. To reduce the number of subsurface atoms, the cluster was truncated into a conical shape, which becomes narrower in lower-lying layers. In the literature, no strong dependence of the XPS calculations on the specific cluster termination was found~\cite{Kahk2019}, which we can confirm.
Still, the choice of both cluster models for the molecular crystal and the surface system is somewhat arbitrary, forming a draw-back of this method.

\Fref{fig:azulene_system_comparison}a shows a comparison of the XPS calculation results for both approaches and all three systems. The overall spectra were shifted (to account for the difference in absolute binding energies due to the charge background issue within the PBCs) and a broadening scheme was applied to simulate experimental resolution. Both methods agree very well for all systems, be it the free molecule, the molecular crystal or the molecule adsorbed on the metal surface. This agreement is a very important result, because it proves that cluster and periodic approaches can yield compatible results, despite their respective challenges. For the systems investigated here, both approaches appear to provide excellent agreement for the obtained relative binding energies. We note, however, that the cluster calculations performed with FHI-aims, both for the molecular crystal and the molecule adsorbed on the surface were much more computationally demanding then the periodic calculations. The all-electron calculations required vastly more computational resources and proved to be less stable with regards to electronic convergence than the periodic calculations, which are straightforward as implemented in CASTEP.

The calculated spectra also provide insight into the interactions present in each system. It is apparent that the interaction between the azulene molecules in the molecular crystal exerts only a minor influence on the XPS peak, which retains its gas-phase shape. The interaction between the molecule and the Cu(111) surface, on the other hand, changes the peak shape a lot, which is an indication of the increased molecule-metal interaction and was already discussed in connection to the experimental data elsewhere~\cite{Klein2019,Klein2019a}. For the adsorbed azulene, we further note that coverage as modelled within PBCs is different to the low-coverage situation captured by the cluster. However, the coverage only exerts a minor influence on the XPS binding energies for this system, as was also observed in experiment.

\begin{figure}
    \centering
    \includegraphics[width=\linewidth]{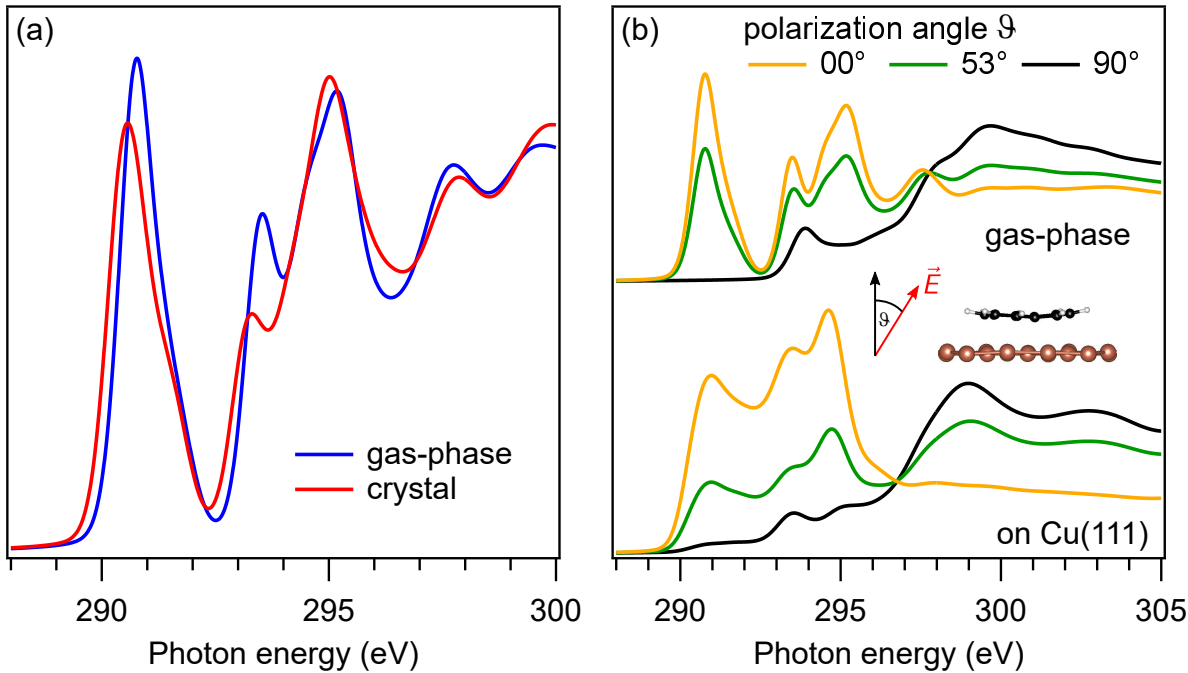}
    \caption{NEXAFS simulations of azulene in the periodic systems with the plane-wave code CASTEP (a) Comparison of the spectra of the gas-phase molecule and the molecular crystal. (b) Comparison of the spectra of the gas-phase molecule and the molecule adsorbed on Cu(111) including angular dependence on the polarization direction of the incident radiation relative to the surface or molecular plane, respectively. The NEXAFS spectra include broadening to simulate a typical experimental resolution.}
    \label{fig:azulene_system_comparison_NEXAFS}
\end{figure}

The NEXAFS spectra of all three systems, calculated with the $\Delta$IP-TP approach in the plane-wave code CASTEP, are compared in \fref{fig:azulene_system_comparison_NEXAFS}. The spectra were not shifted with respect to each other. The good agreement, despite the presence of the space-charge effect, is due to the ionization potential correction ($\Delta$IP) of the NEXAFS transitions according to the C1s binding energy of each carbon. In \fref{fig:azulene_system_comparison_NEXAFS}a it can be seen that the spectra of free molecule and molecular crystal are very similar. For the NEXAFS spectrum of the adsorbed molecule (\fref{fig:azulene_system_comparison_NEXAFS}b), the angular dependence of the NEXAFS transitions with regard to the polarization vector is important. Therefore we chose to simulate three different polar angles~$\vartheta$ for the polarization direction. Two of those correspond to the polarization direction parallel (90$^\circ$) and orthogonal to the surface (0$^\circ$), while the third is the so called \emph{magic angle} (53$^\circ$). The \emph{magic angle} is the polarization angle, at which the orientation of the transition dipole moment $\boldsymbol{\mu}_{fi}$ relative to the substrate doesn't influence the transition intensity.  It should be noted that the ideal magic angle (for our threefold-symmetric substrate) is in fact 54.7$^\circ$~\cite{Stoehr1992}. However, the actual magic angle encountered in an experiment is dependent on the degree of polarization of the used radiation~\cite{Stoehr1992,Breuer2015}. The value for the degree of polarization of 0.9 used in the corresponding experiments~\cite{Klein2019} leads to a magic angle of 53$^\circ$. For all polar angles, the spectra are averaged over the azimuthal angle, due to the symmetry of the substrate (see \eref{eq:polar_ave}). The comparison between NEXAFS of gas-phase and adsorbed molecule reveals stark differences in the spectra, giving insight into the mechanism of the molecule-surface interaction. While the overall dichroism reveals that the molecule is approximately flat on the Cu(111) surface, there is another telling detail in the spectra. The non-vanishing intensity of the C 1s to LUMO transition (291 eV) for the 90$^\circ$ polar angle shows that the clear separation of $\sigma^*$ and $\pi^*$ states (as present for the free molecule) is now broken due to the hybridization between the electronic states of adsorbed molecule and metal surface~\cite{Mainka1995,Diller2017}. 
The corresponding experimental spectra, the simulated data, and the subsequent interpretation were already reported in previous publications and show a good agreement between experiment and theory~\cite{Klein2019,Klein2019a}.

\subsection{\label{section-charge} The effect of charge in periodic core-hole simulations}

We use this opportunity of having reliable experimental data for azulene thin films and azulene monolayers adsorbed on the Cu(111) surface to show a systematic comparison of the $\Delta$IP-TP method with the similar $\Delta$IP-XTP calculations. The advantage of the XTP method is that it creates a net-neutral unit cell which does not require a spurious uniform background counter charge, which is otherwise present in the periodic $\Delta$SCF and ($\Delta$IP-)TP calculations and can significantly influence electrostatic properties~\cite{Taucher2020}. The array of calculations making up a $\Delta$IP-XTP simulation differs only in the third step (as shown in \fref{fig:workflow}c) from the $\Delta$IP-TP method: We perform $N$ half core-hole (but zero charge) XTP calculations where we introduce half an electron into the lowest unoccupied state to compensate for the core-hole.

\begin{figure}
    \centering
    \includegraphics[width=\linewidth]{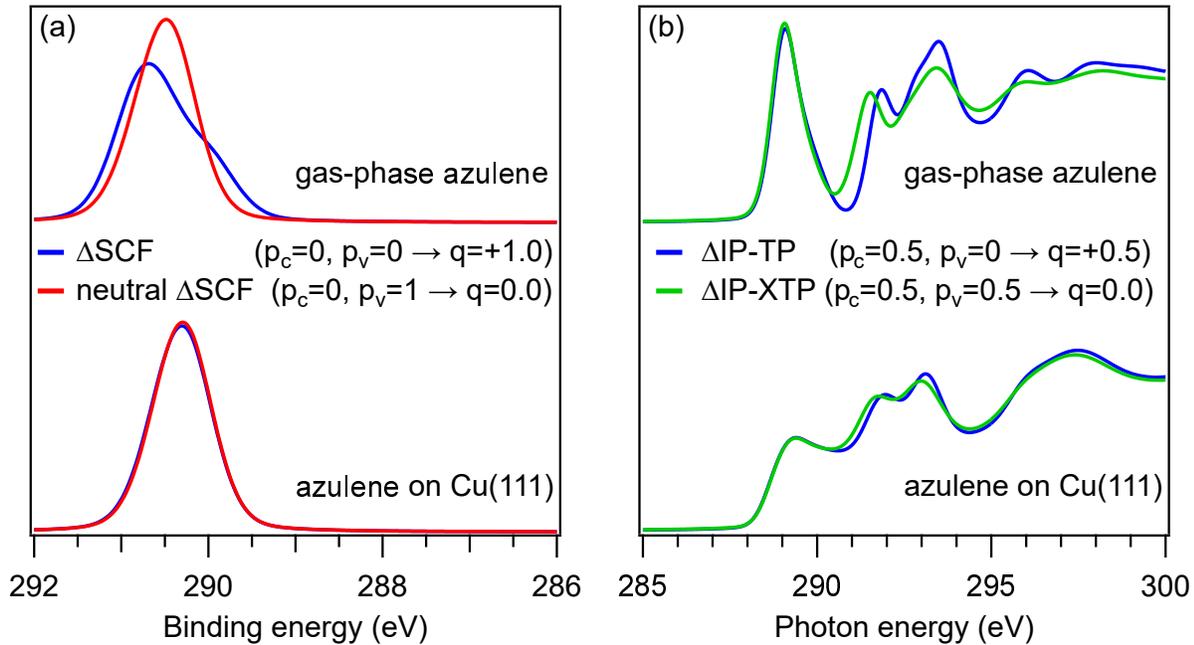}
    \caption{Core-level spectra calculated with and without background charge  compensation for azulene molecule in the gas-phase and adsorbed on the Cu(111) surface. (a) Comparison of $\Delta$SCF XP spectra calculations with a homogeneous compensation charge (blue) and with a compensation charge introduced in the lowest unoccupied state (red). (b) Comparison of the $\Delta$IP-TP (blue) and $\Delta$IP-XTP (green) NEXAFS spectra. Both spectra share a 0.5 e$^-$ core-hole. The TP calculation includes -0.5 e background charge (blue) while the XTP calculations compensate charge by adding half an electron into the lowest unoccupied state (green). All spectra include broadening to simulate a typical experimental resolution and were shifted for better comparison.}
    \label{fig:XTP_vs_TP}
\end{figure}

The results of the $\Delta$IP-TP and $\Delta$IP-XTP simulations are compared for the azulene molecule in the gas phase and adsorbed on the copper surface in \fref{fig:XTP_vs_TP}. In this figure all spectra were rigidly shifted to compensate for the global shift due to the different charge state of the systems and therefore enable a better inspection of their spectral shape.
Both methods produce virtually indistinguishable XPS peak shapes for the molecule adsorbed on the metal surface, while stark differences are present for the gas-phase molecule (see \fref{fig:XTP_vs_TP}a). This fact is unsurprising, because the peculiar peak shape of the gas-phase azulene molecule is due to the localized charge distribution, as already discussed in the literature on the backdrop of experimental XPS data for thin multicrystalline films of azulene~\cite{Klein2019}. 
The neutral $\Delta$SCF calculations do not remove the electron taken from the core-state, but instead promote it into the lowest unoccupied state. The occupation of the former LUMO has massive consequences for the electronic structure of the gas-phase molecule and therefore the resulting spectra are negatively influenced. On the other hand, in the case of azulene adsorbed on Cu(111), the continuous density of states of the metal surface around the Fermi level has the effect that the additional electron does not present a significant change in the electronic structure and the resulting spectra remain similar.

For the proper simulation of the NEXAFS spectra in \fref{fig:XTP_vs_TP}b,  the obvious error in the neutralized $\Delta$SCF calculations for the free molecule made it necessary to perform the $\Delta$IP shift both for the TP and XTP simulation on the basis of the charged $\Delta$SCF calculations. The results show that again the spectra for the molecule adsorbed on the metal surface are almost identical for TP and XTP, while there are some differences for the gas-phase molecule.

In comparison to various core-hole constraining methods, Michelitsch and Reuter have previously reported that the XTP method (note that no $\Delta$IP correction is applied) on balance provides the best agreement with experiment for absolute energies and intensities of first and second transition energies of NEXAFS spectra for a range of isolated organic molecules~\cite{Michelitsch2019}. We also compared  the $\Delta$IP-TP and $\Delta$IP-XTP spectra with the experimental data for the multicrystalline azulene films from reference~\cite{Klein2020} (see figure S6 in the supplemental material) and found a similar performance for TP and XTP. In the future we will provide a more systematic study of method differences on a larger set of benchmark systems, to test if the performance of the $\Delta$IP-TP and $\Delta$IP-XTP approaches is robust across a variety of cases including more challenging systems, e.g. containing coherent dipolar arrangements~\cite{Taucher2020}.

\section{Conclusion and Future Outlook}

We have presented the numerical and technical details of Density-Functional-Theory-based core-hole simulation methods to calculate 1s XPS and K-edge NEXAFS signatures of organic molecules, organic molecular crystals, and metal-organic interfaces. The introduced methods are variants of the $\Delta$SCF method and the transition potential approach ($\Delta$IP-TP). For the benefit of practitioners in the field, we have compared and contrasted key parameter choices in performing such simulations including the choice of exchange-correlation functional, basis set,  and if the model is to be aperiodic or treated within periodic boundary conditions. We find that all tested functionals perform well for absolute and relative XPS binding energies, as long as the shifts are not too extreme due to strong descreening. For such extreme shifts, as encountered in the ETFA molecule, hybrid functionals perform better than GGAs and meta-GGAs. For periodic calculations, the calculation of absolute energies is more difficult, however, we find that periodic and aperiodic calculations yield comparable results for relative shifts, even in the presence of the inherent space charge problem due to the use of a compensating background charge.

The choice of periodic or aperiodic description for the simulation of XPS and NEXAFS spectra is strongly dependent on the system under investigation and comes with a complex trade-off regarding advantages and disadvantages. For example, aperiodic atomic-orbital based calculations, as we have performed with the FHI-aims package, can provide accurate core-level binding energies~\cite{Kahk2019} and NEXAFS transitions~\cite{Michelitsch2019} for an isolated molecule, a molecular crystal, and a metal-adsorbed molecule, but lack robustness with respect to core-hole localization and provide substantial computational overhead for the latter two systems. More advanced approaches to core-hole localization based on excited-state orbital optimization~\cite{Hait2020} or more sophisticated constraint definition~\cite{Ferre2002,Loos2007,Maurer2013} may be necessary to resolve this generally. In the case of a periodic pseudopotential plane wave description, as provided within CASTEP, the core-hole localization is in-built and the computations are efficient and numerically robust. However, the employed periodic boundary conditions can introduce an artificial charge background and spurious electrostatics with some core-hole constraining approaches~\cite{Taucher2020}. We show that both, aperiodic and periodic approaches yield virtually identical relative core-level binding energies and that the relative peak positions in XPS and NEXAFS spectra do not appear to be strongly affected, even when different charge compensation schemes are used. However, we note that other systems with strong dipoles and no Fermi level pinning may show more significant artefacts in spectra when a homogeneous charge background is present. Furthermore, the comparison of absolute binding energies between simulation and experimental data for condensed matter systems carries additional pitfalls not addressed in this publication. We will systematically assess both issues in future work.

Despite the recent emergence of many highly accurate first-principles approaches to simulate XPS and XAS signatures based on many-body perturbation theory~\cite{Gilmore2015,Liang2017,Liang2018,Liang2019}, approximate core-hole simulation methods remain essential workhorses for highly complex spectroscopy simulations with applications ranging from in-operando electrocatalysis~\cite{Nattino2020} to ultrafast dynamics~\cite{Northey2020} and pump-probe spectroscopy~\cite{Ehlert2018}. The efficiency of these methods  enables the creation of high-volume data~\cite{Mathew2018,Aarva2019,Pi2020} which can be used for machine-learning-assisted spectral assignment~\cite{Aarva2019a,Zheng2018,Rankine2020}.
The here presented $\Delta$SCF and $\Delta$IP-TP approaches as implemented in CASTEP are versatile and reliable for organic molecules in different environments and represent important tools in this context.

\section{Conflict of Interest}
The authors declare no conflict of interest.

\section{Author Information}
\subsection{Corresponding Author}
Reinhard J. Maurer, r.maurer@warwick.ac.uk 
\subsection{ORCIDs}
Benedikt P. Klein: 0000-0002-6205-8879\\
Samuel J. Hall: 0000-0003-3765-828X \\
Reinhard J. Maurer: 0000-0002-3004-785X

\section{Supporting information}

\ack

S.J.H and R.J.M. acknowledge funding for a PhD studentship through the EPSRC Centre for Doctoral Training in Molecular Analytical Science (EP/L015307/1) and computing resources via the EPSRC-funded HPC Midlands+ computing centre (EP/P020232/1) and the EPSRC-funded Materials Chemistry Consortium for the ARCHER UK National Supercomputing Service (EP/R029431/1). R.J.M acknowledges support via a UKRI Future Leaders Fellowship (MR/S016023/1). The authors acknowledge many fruitful and enjoyable discussions with Katharina Diller and David Duncan.

\printbibliography

\end{document}

\pdfminorversion=4

\title[SI: Ab-Initio Core-Hole Simulations]{Supplementary Material: \\ 
~ \\
~ \\
The Nuts and Bolts of Ab-Initio Core-Hole Simulations for K-shell X-Ray Photoemission and Absorption spectra}

\author{Benedikt P Klein$^{1,2}$, Samuel J Hall$^{1,3}$ and Reinhard J Maurer$^1$}
\address{$^1$Department of Chemistry, University of Warwick, Gibbet Hill Rd, Coventry, CV4 7AL, UK}
\address{$^2$Diamond Light Source, Harwell Science and Innovation Campus, Didcot, OX11 0DE, United Kingdom}
\address{$^3$MAS CDT, Senate House, University of Warwick, Gibbet Hill Rd, Coventry, CV4 7AL, UK }

\ead{r.maurer@warwick.ac.uk}
\vspace{10pt}
\begin{indented}
\item[]\today
\end{indented}

\maketitle
%%%%%%%%%%%%%%%%%%%%%%%% here it is %%%%%%%%%%%%%%%%%%%%%%%%%%%%%%%%%
%%%%%%%%%%%%%%%%%%%%%%%% here it is %%%%%%%%%%%%%%%%%%%%%%%%%%%%%%%%%

\tableofcontents

\clearpage
\markboth{}{}

\section{Performance of the XC functional for the calculation of XPS binding energies}

\begin{table}[h]
\caption{Comparison of the absolute carbon 1s binding energies of ETFA calculated  with various XC functionals with the experimental values~\cite{Gelius1973}. All theoretical values were calculated using the $\Delta$SCF approach in the all-electron code FHI-aims with a core-augmented "tight-tier2" basis set. All energies in eV.}
\begin{center}
\footnotesize
\lineup
\begin{tabular}{l|ccccccccc}
\br
Carbon & Expt.  & PBE    & PW91   & HES06  & PBE0   & B3LYP  & SCAN   & TPSS   & xDH-PBE0 \\\mr
C1      & 299.45 & 297.52 & 297.88 & 298.44 & 298.39 & 299.00 & 298.78 & 298.35 & 298.39   \\
C2      & 296.01 & 294.52 & 294.88 & 295.43 & 295.38 & 295.99 & 295.72 & 295.32 & 295.38   \\
C3      & 293.07 & 292.32 & 292.64 & 292.85 & 292.79 & 293.38 & 293.24 & 293.09 & 292.79   \\
C4      & 291.20 & 290.79 & 291.11 & 291.21 & 291.15 & 291.76 & 291.64 & 291.58 & 291.15  \\\br
\end{tabular}
\end{center}
\label{tab:ETFA_abs}
\end{table}

\begin{table}[h]
\caption{Comparison of the relative shifts in the carbon 1s binding energies of ETFA calculated  with various XC functionals with the experimental values~\cite{Gelius1973}. All theoretical values were calculated using the $\Delta$SCF approach in the all-electron code FHI-aims with a core-augmented "tight-tier2" basis set. All shifts are in eV and were calculated with respect to C4 using the values from \tref{tab:ETFA_abs}.}
\begin{center}
\footnotesize
\lineup
\begin{tabular}{l|ccccccccc}
\br
Carbon & Expt.  & PBE    & PW91   & HES06  & PBE0   & B3LYP  & SCAN   & TPSS   & xDH-PBE0 \\\mr
C1 & 8.25 & 6.73 & 6.77 & 7.23 & 7.24 & 7.24 & 7.14 & 6.77 & 7.24 \\
C2 & 4.81 & 3.73 & 3.77 & 4.22 & 4.23 & 4.23 & 4.08 & 3.74 & 4.23 \\
C3 & 1.87 & 1.52 & 1.53 & 1.64 & 1.64 & 1.62 & 1.60 & 1.51 & 1.64 \\
C4 & 0.00 & 0.00 & 0.00 & 0.00 & 0.00 & 0.00 & 0.00 & 0.00 & 0.00 \\\br
\end{tabular}
\end{center}
\label{tab:ETFA_rel}
\end{table}

\begin{table*}[h]
\caption{Absolute carbon 1s binding energies of the azupyrene molecule in dependence of the XC functional, calculated using the $\Delta$SCF approach in the all-electron code FHI-aims with a core-augmented "tight-tier2" basis set. All shifts are in eV.}
\footnotesize
\lineup
\begin{center}
\begin{tabular}{l|cccccccc}
\br
Carbon & PBE & PW91 & HSE06 & PBE0 & B3LYP & SCAN & TPSS & xDH-PBE0 \\\mr
C1 & 288.90 & 289.20 & 289.31 & 289.26 & 289.78 & 289.61 & 289.56 & 289.26 \\
C2 & 289.14 & 289.45 & 289.62 & 289.58 & 290.08 & 289.87 & 289.81 & 289.58 \\
C3 & 289.15 & 289.45 & 289.57 & 289.52 & 290.00 & 289.84 & 289.75 & 289.52 \\
C4 & 288.71 & 289.02 & 289.13 & 289.08 & 289.58 & 289.42 & 289.37 & 289.08 \\
C5 & 289.72 & 290.02 & 290.22 & 290.17 & 290.63 & 290.47 & 290.32 & 290.17\\\br
\end{tabular}
\label{tab:azupyrene_abs}
\end{center}
\end{table*}

\begin{table*}[h]
\caption{Relative shifts of the carbon 1s binding energies of the azupyrene molecule in dependence of the XC functional, calculated according to Koopmans' theorem as $E_\mathrm{B}(i) = -\varepsilon_{i,\mathrm{KS}}$ in the all-electron code FHI-aims with a core-augmented "tight-tier2" basis set. The energies belonging to different KS states localized on the same symmetry inequivalent carbon atoms were averaged. All shifts are in eV and w.r.t. the binding energy of C1.}

\begin{center}
\begin{tabular}{l|cccccccc}
\br
Carbon & PBE & PW91 & HSE06 & PBE0 & B3LYP & SCAN & TPSS & XDH-PBE0 \\\mr
C1 & 0.00 & 0.00  & 0.00  & 0.00  & 0.00  & 0.00  & 0.00  & 0.00  \\
C2 & 0.30  & 0.30  & 0.37  & 0.38  & 0.36  & 0.36  & 0.31  & 0.38  \\
C3 & 0.24 & 0.24  & 0.25  & 0.25  & 0.22  & 0.24  & 0.19  & 0.25  \\
C4 & -0.14 & -0.14 & -0.15 & -0.15 & -0.16 & -0.15 & -0.14 & -0.15 \\
C5 & 0.81 & 0.81  & 0.88  & 0.88  & 0.84  & 0.85  & 0.77  & 0.88    \\   \br
\end{tabular}
\label{tab:azupyrene_koopmans}
\end{center}
\end{table*}

%\begin{figure}[h!]
%    \centering
%    \includegraphics[width=0.5\linewidth]{_SI_01_XC_ETFA_relative.eps}
%    \caption{Comparison of the C 1s XPS spectra of ETFA calculated with different XC functionals on a relative binding energy scale w.r.t. C4. Upper part: experimental spectrum from Ref. \cite{Gelius1973} (open circles, data points; red line, fit function). Lower part: calculated spectra. Vertical lines as guide-for-the-eye. All spectra were obtained by applying a pseudo-Voigt broadening on the calculated C 1s binding energies from table~\ref{tab:ETFA_abs} to mimic a typical experimental resolution.}
%    \label{fig:SI_XC_ETFA}
%\end{figure}

\clearpage

\section{Basis set convergence of $\Delta$SCF calculations with FHI-aims}

\begin{table}[h!]
\centering
\caption{Absolute binding energies for azupyrene, calculated with PBE and the \emph{force\_occupation\_basis} (FOB) method in FHI-aims, including core-augmentation functions.}
\begin{tabular}{l|ccc|ccc|ccc}
\br
&  \multicolumn{3}{c}{light}  & \multicolumn{3}{c}{tight}   &  \multicolumn{3}{c}{really-tight} \\\mr
Atom & tier1  & tier2  & tier3  & tier1  & tier2  & tier3  & tier1  & tier2        & tier3  \\
1    & 289.13 & 288.88 & 288.79 & 289.15 & 288.90 & 288.81 & 289.15 & 288.90       & 288.81 \\
2    & 289.36 & 289.12 & 289.04 & 289.39 & 289.14 & 289.06 & 289.39 & 289.14       & 289.06 \\
3    & 289.32 & 289.13 & 289.06 & 289.34 & 289.15 & 289.08 & 289.34 & 289.15       & 289.08 \\
4    & 288.96 & 288.69 & 288.61 & 288.98 & 288.71 & 288.63 & 288.98 & 288.71       & 288.63 \\
5    & 289.87 & 289.70 & 289.63 & 289.90 & 289.72 & 289.65 & 289.90 & 289.72       & 289.65 \\
\br
\end{tabular}
\label{tab:basis_convergence_azpyr_abs}
\end{table}

\begin{table}[h!]
\centering
\caption{Relative binding energies for azupyrene, calculated with PBE and the FOB method in FHI-aims, including core-augmentation functions.}
\begin{tabular}{l|ccc|ccc|ccc}
\br
 &    \multicolumn{3}{c}{light}    &  \multicolumn{3}{c}{tight} &  \multicolumn{3}{c}{really-tight} \\\mr
Atom & tier1 & tier2 & tier3 & tier1 & tier2 & tier3 & tier1 & tier2        & tier3 \\
1    & 0.00  & 0.00  & 0.00  & 0.00  & 0.00  & 0.00  & 0.00  & 0.00         & 0.00  \\
2    & 0.23  & 0.25  & 0.25  & 0.23  & 0.25  & 0.25  & 0.23  & 0.25         & 0.25  \\
3    & 0.18  & 0.25  & 0.27  & 0.19  & 0.25  & 0.27  & 0.19  & 0.25         & 0.27  \\
4    & -0.18 & -0.19 & -0.19 & -0.17 & -0.18 & -0.18 & -0.17 & -0.18        & -0.18 \\
5    & 0.74  & 0.82  & 0.84  & 0.75  & 0.82  & 0.84  & 0.75  & 0.82         & 0.84 \\
\br
\end{tabular}
\label{tab:basis_convergence_azpyr_rel}
\end{table}

\begin{table}[h!]
\centering
\caption{Absolute binding energies for ETFA, calculated with PBE and the FOB method in FHI-aims, including core-augmentation functions.}
\begin{tabular}{l|cc|cc|cc}
\br
    &    \multicolumn{2}{c}{light}  &   \multicolumn{2}{c}{tight}  &  \multicolumn{2}{c}{really-tight} \\\mr
Atom & tier1  & tier2  & tier1  & tier2  & tier1  & tier2        \\
1    & 297.80 & 297.51 & 297.79 & 297.52 & 297.79 & 297.53       \\
2    & 294.80 & 294.51 & 294.82 & 294.52 & 294.82 & 294.52       \\
3    & 292.55 & 292.31 & 292.58 & 292.32 & 292.58 & 292.32       \\
4    & 291.05 & 290.78 & 291.07 & 290.79 & 291.07 & 290.79      \\
\br
\end{tabular}
\label{tab:basis_convergence_ETFA_abs}
\end{table}

\begin{table}[h!]
\centering
\caption{Relative shifts in the binding energies for ETFA, calculated with PBE and the FOB method in FHI-aims, including core-augmentation functions}
\begin{tabular}{l|cc|cc|cc}
\br
    &    \multicolumn{2}{c}{light}  &   \multicolumn{2}{c}{tight}  &  \multicolumn{2}{c}{really-tight} \\
    \mr
Atom & tier1 & tier2 & tier1 & tier2 & tier1 & tier2        \\
1    & 0.00  & 0.00  & 0.00  & 0.00  & 0.00  & 0.00         \\
2    & -2.99 & -3.00 & -2.98 & -3.00 & -2.98 & -3.00        \\
3    & -5.24 & -5.21 & -5.22 & -5.21 & -5.22 & -5.21        \\
4    & -6.75 & -6.73 & -6.72 & -6.73 & -6.72 & -6.73       \\
\br
\end{tabular}
\label{tab:basis_convergence_ETFA_rel}
\end{table}

\clearpage

\section{Influence of the pseudopotential projectors on XPS and NEXAFS spectra for azupyrene}

\begin{figure}[h]
    \centering
    \includegraphics[width=\linewidth]{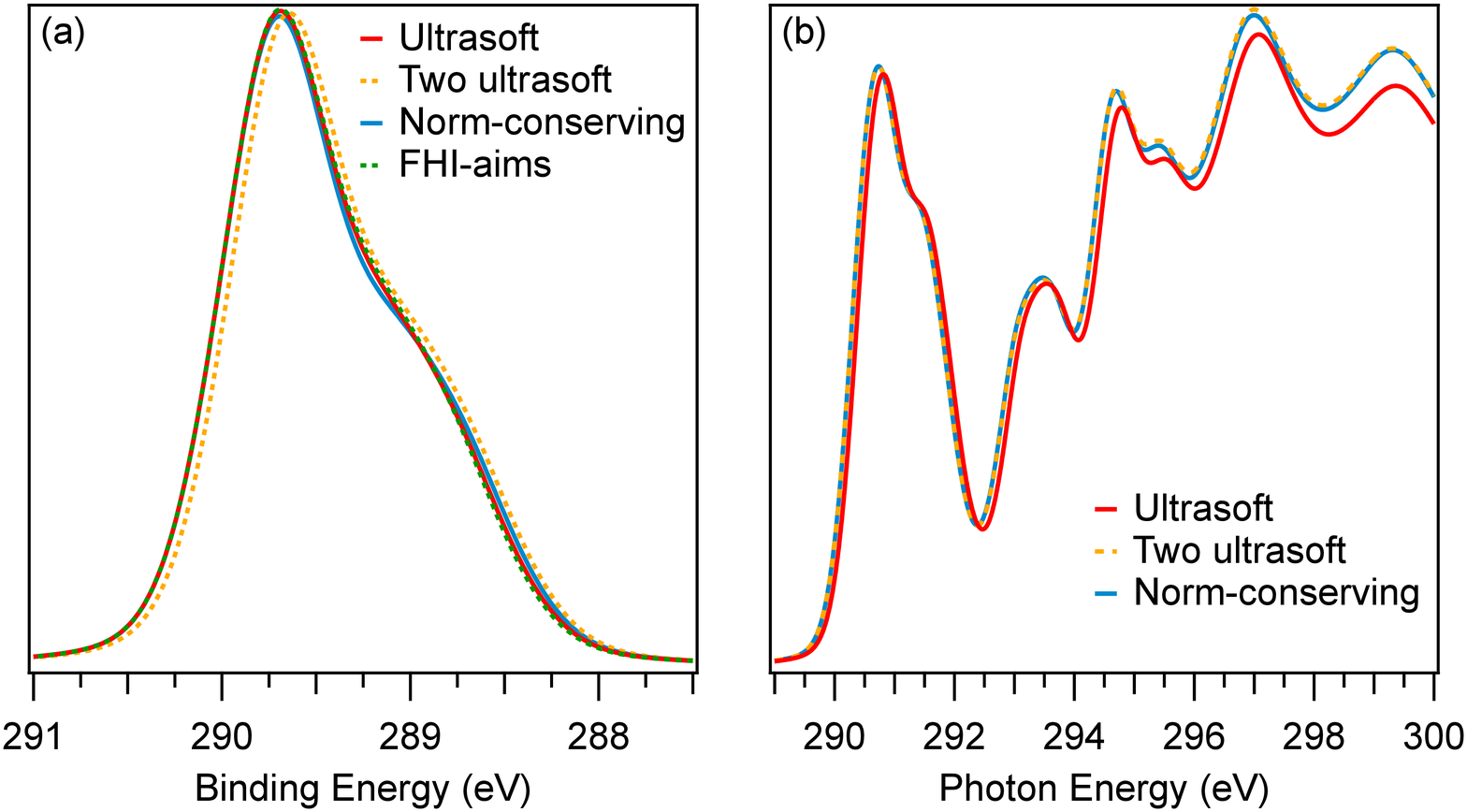}
    \caption{Comparison of the XPS spectra of gas-phase azupyrene (a), obtained when using three different on-the-fly generated pseudopotential projectors: red, ultrasoft projectors; blue, two ultrasoft projectors per angular momentum channel; green, norm-conserving projectors and FHI-aims FOB spectra in orange. All three CASTEP spectra were shifted by -6.1 eV to align up with FHI-aims spectra. (b) shows total NEXAFS spectra of all three different pseudopotentials.}
    \label{fig:pseudopotentials_converge}
\end{figure}

\begin{table}[h!]
\centering
\caption{Absolute XPS binding energies for azupyrene calculated (a) in the all-electron code FHI-aims with a core-augmented "tight-tier2" basis set and the FOB approach and (b) with CASTEP and the three different pseudopotential projectors. All calculations were performed with the PBE functional.}
\begin{tabular}{c|c|c|c|c}
\br
Atom & FHI-aims & Ultrasoft & Two-Ultrasoft & Norm-Conserving \\ \mr
1    & 288.90   & 294.98    & 294.92        & 294.94 \\    
2    & 289.14   & 295.23    & 295.17        & 295.19 \\
3    & 289.15   & 295.25    & 295.19        & 295.24 \\
4    & 288.71   & 294.79    & 294.73        & 294.76 \\
5    & 289.72   & 295.82    & 295.77        & 295.82 \\
\br
\end{tabular}
\label{tab:azpyr_rel_pseudopot}
\end{table}

\clearpage

\section{Convergence behaviour  of NEXAFS spectra for azupyrene}

\begin{figure}[h!]
    \centering
    \includegraphics[width=\linewidth]{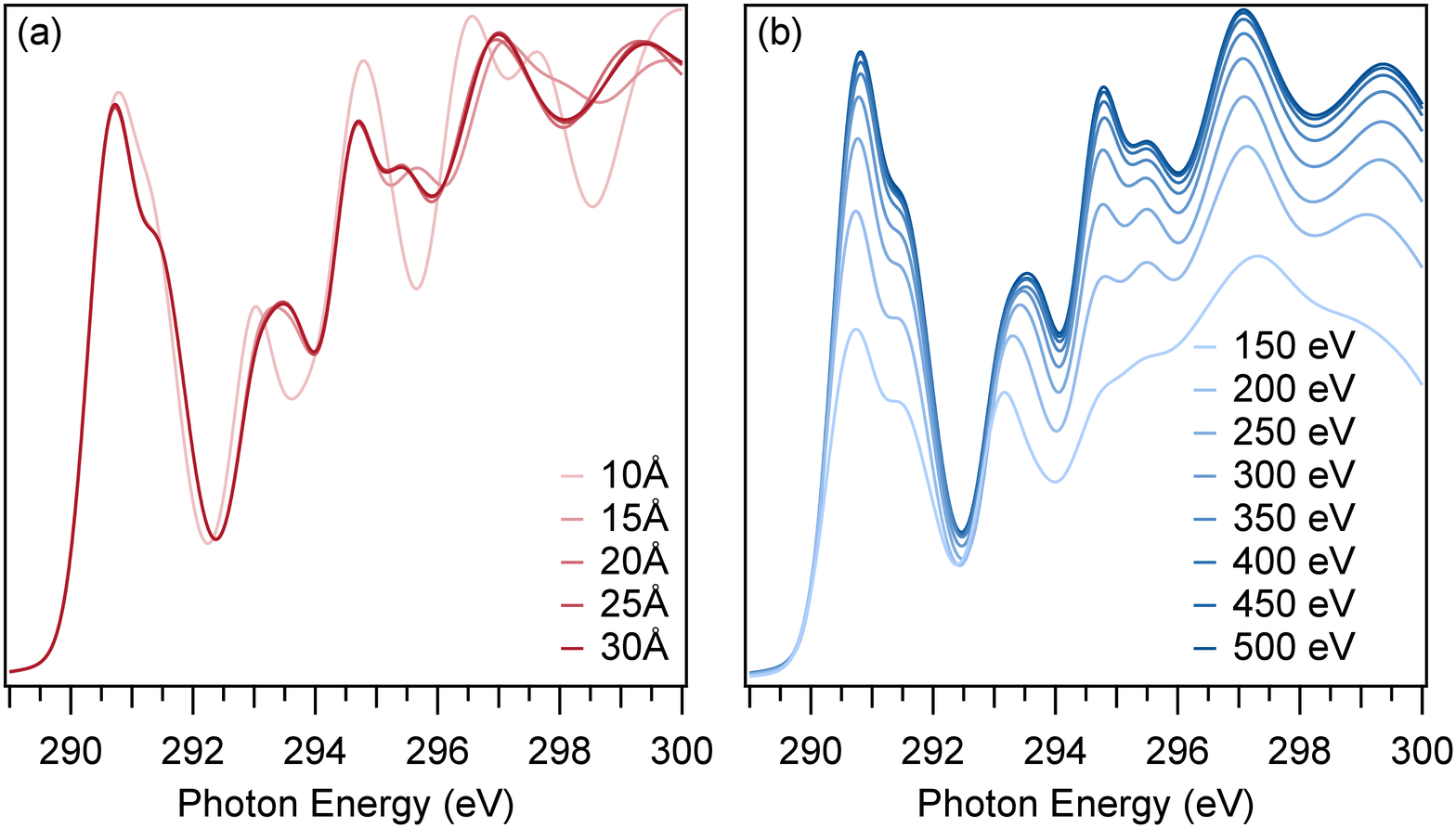}
    \caption{ Convergence of the NEXAFS spectrum of azupyrene with respect to increasing box size (a) and cut-off energy of the plane wave basis (b). A constant cut-off energy of 400 eV was used in the box size convergence and a 30 \AA{} box size used for cut-off convergence. All spectra were calculated with ultrasoft pseudopotentials. Spectra in (a) have been shifted to align leading edges to the same energy.}
    \label{fig:SI_NEXAFS_convergence}
\end{figure}

\clearpage

\section{Influence of the XC functional on the NEXAFS spectrum for azupyrene}

\begin{figure}[h!]
    \centering
    \includegraphics[width=0.5\linewidth]{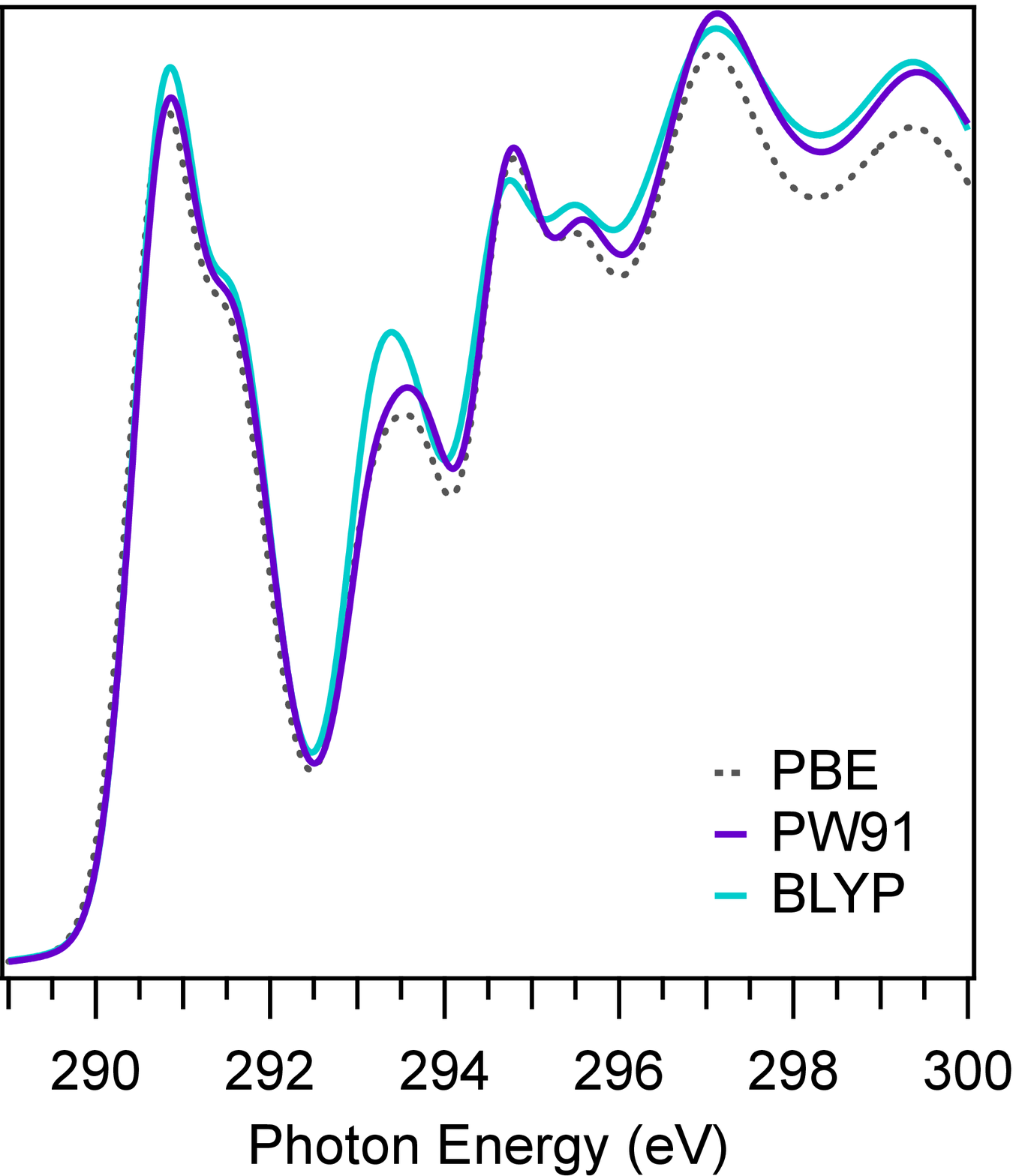}
    \caption{Comparison of NEXAFS spectra calculated with three different GGA XC functionals: PBE (grey dashes), PW91 (purple) and BLYP (cyan). All spectra were calculated with ultrasoft pseudopotentials, and a cut-off energy of 500 eV.}
    \label{fig:Si_XC_NEXAFS}
\end{figure}

\clearpage

\section{Convergence of XPS and NEXAFS spectra for azulene molecular crystal and azulene adsorbed on Cu(111)}

\begin{figure}[h]
    \centering
    \includegraphics[width=\linewidth]{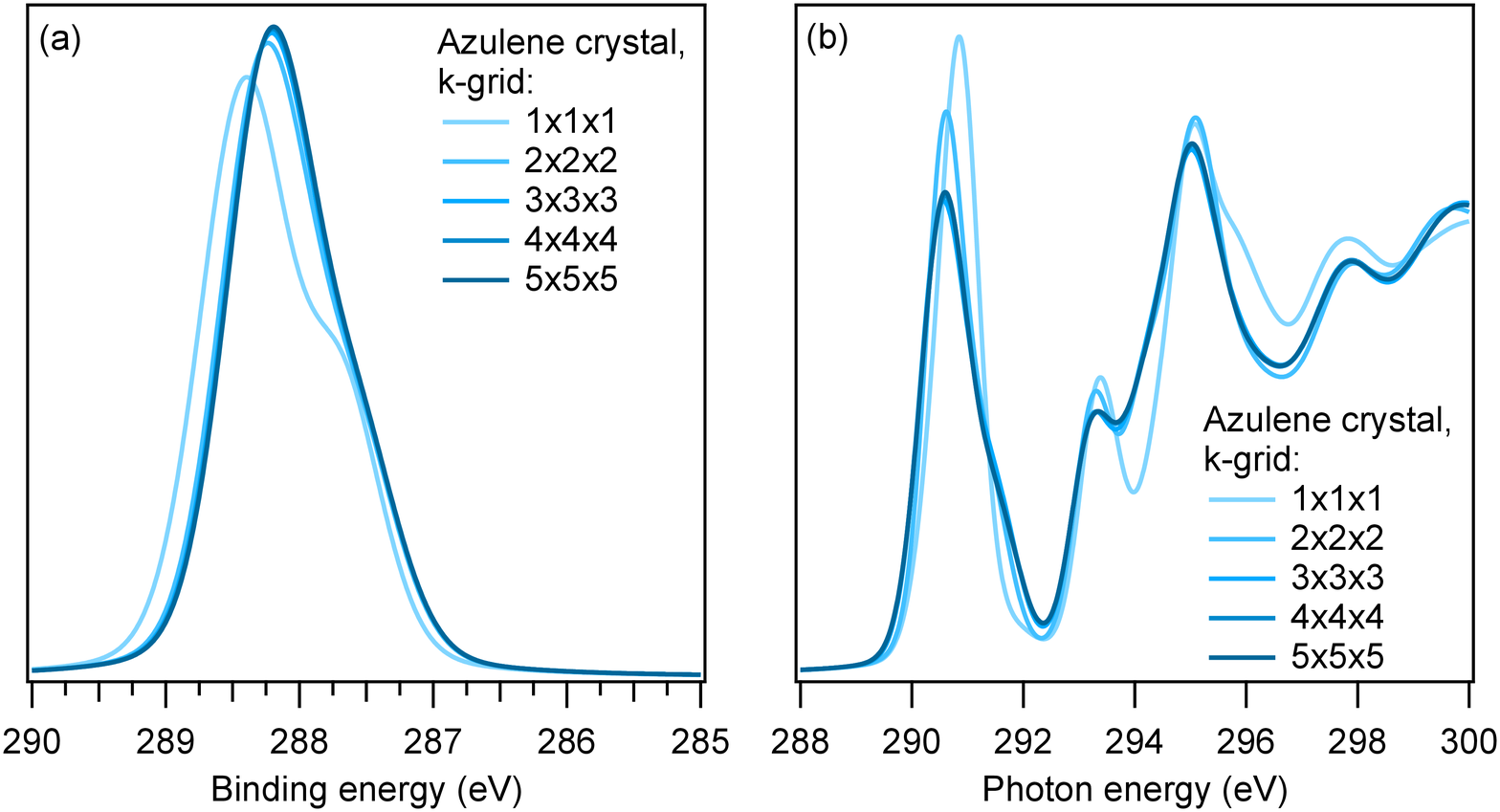}
    \caption{K-space convergence series for the molecular azulene crystal. (a) XPS spectra, (b) NEXAFS spectra. The system is a molecular crystal with four molecules per unit cell.}
    \label{fig:k_convergence_azulene_crystal}
\end{figure}

One additional issue of condensed systems compared to gas-phase calculations is the sampling of electronic states across the periodic crystal via Brillouin zone integration (or k-space sampling). Therefore, the convergence of both the XPS and NEXAFS calculations has to be ensured with respect to the number of k points included. \Fref{fig:k_convergence_azulene_crystal} shows the k-convergence both of the XPS and NEXAFS spectra of the azulene molecular crystal as a function of k points in all directions in a regular Monkhorst-Pack k-grid~\cite{Monkhorst1976}. The XPS spectra in \fref{fig:k_convergence_azulene_crystal}a are already converged for a 2$\times$2$\times$2 k-grid, while the NEXAFS spectra in \fref{fig:k_convergence_azulene_crystal}b are converged for a 3$\times$3$\times$3 k-grid. In both cases there is a significant difference in the spectra, if the k-grid is not chosen large enough. The required k-grids correspond to k-point densities similar to what is required to yield converged bulk cohesive energies of molecular crystals.

For azulene adsorbed on Cu(111), we simulate the system in a slab approach with 3-dimensional PBCs. In this approach, we model the slab as a 2D extended surface with a vacuum layer on top that separates the surface from its periodic image perpendicular to the surface. In a system represented by a slab approach, the k-grid convergence in the 2D periodic plane has to be tested, as well as the convergence with respect to the thickness of the vacuum layer. In addition, the convergence with respect to cut-off energy has to be revisited, because now surface atoms with different pseudopotentials are present. \Fref{fig:convergence_azulene_cu} shows the convergence behaviour of the NEXAFS spectra for all three parameters. Regarding the k-grid, the NEXAFS spectrum is already quite quickly converged at the 3$\times$3$\times$1 k-grid (\fref{fig:convergence_azulene_cu}a). This is surprising as even the convergence of the adsorption energy requires a larger k-grid for this system~\cite{Klein2019}. This finding indicates that the electronic states involved in the NEXAFS transitions of this system are probably quite localized on the adsorbate and more strongly interacting metal-organic interfaces will show more metal contributions and require denser k-grids. The convergence with respect to plane-wave cut-off energy is similar to what is found for the gas-phase molecule (\fref{fig:convergence_azulene_cu}b). The  thickness of the vacuum slab only shows a minor influence on the NEXAFS spectrum (\fref{fig:convergence_azulene_cu}c).

\begin{figure}[h]
    \centering
    \includegraphics[width=\linewidth]{_SI_06_surface_convergence.eps}
    \caption{Convergence series for azulene adsorbed on Cu(111). (a) k-space convergence with 450 eV cut-off energy, (b) plane-wave cut-off convergence with a constant k-grid of 8$\times$8$\times$1. (c) vacuum slab convergence with 450 eV cut-off energy and a 6$\times$6$\times$1 k-grid.}
    \label{fig:convergence_azulene_cu}
\end{figure}

\clearpage

\begin{figure}
    \centering
    \includegraphics[width=0.5\linewidth]{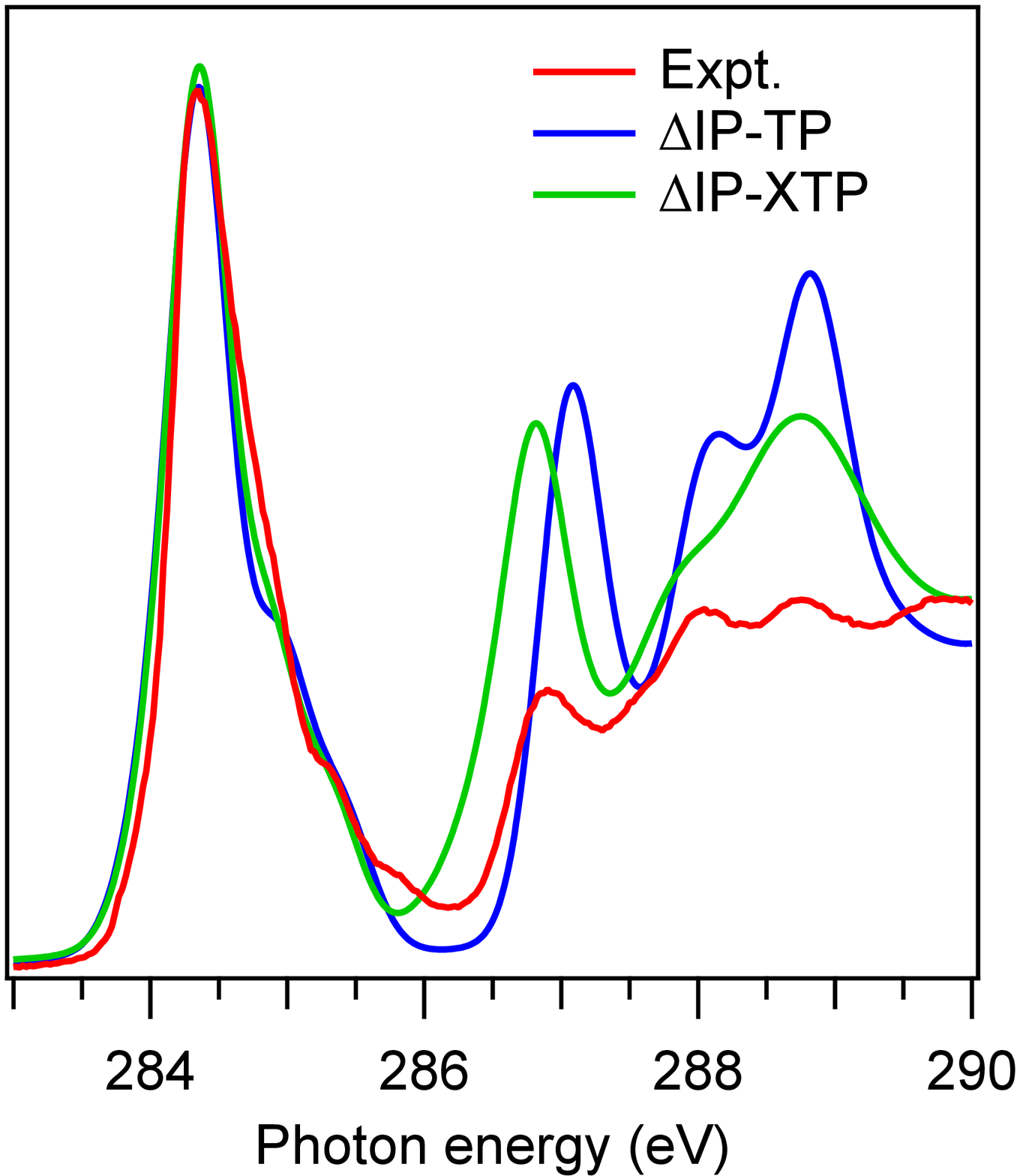}
    \caption{Comparison of the core-level spectra calculated with and without forced charge compensation for azulene molecule in the gas-phase with the experimental data from \cite{Klein2020}. The simulated spectra were shifted by -8.25 eV (XTP) and -6.4 eV (TP) for better comparison of the spectral features. Both XTP and TP simulations include broadening to resemble the experimental resolution. The broadening applied here was directly adjusted according to this specific experimental data and uses slightly different parameters than the "best guess" values mentioned in the main paper. The simulated spectra shown here were obtained by using a FHWM of 0.5~eV combined with a 85\%/15\% G/L ratio in the low energy range, and a 2.0~eV FWHM combined with a 20\%/80\% G/L ratio in the high energy range.}
    \label{fig:XTP_vs_TP}
\end{figure}

\clearpage

\printbibliography